\documentclass[preprint]{aastex62}

\received{February 25, 2020}
\revised{May 31, 2020}
\accepted{June 2, 2020}
\submitjournal{ApJ}

\shorttitle{Chemical Variation among Protostellar Cores}
\shortauthors{Aikawa et al.}

\begin{document}

\title{Chemical Variation among Protostellar Cores: Dependence on Prestellar Core Conditions}

\correspondingauthor{Yuri Aikawa}
\email{aikawa@astron.s.u-tokyo.ac.jp}

\author[0000-0003-3283-6884]{Yuri Aikawa}
\affil{Department of Astronomy, The University of Tokyo,
Tokyo, 113-0033, Japan}

\author{Kenji Furuya}
\affiliation{Center for Computational Sciences, University of Tsukuba,  Tsukuba, 305-8577, Japan}
\affiliation{National Astronomical Observatory of Japan, Tokyo 181-8588, Japan}

\author{Satoshi Yamamoto}
\affiliation{Department of Physics, The University of Tokyo
Tokyo, 113-0033, Japan}

\author{Nami Sakai}
\affiliation{RIKEN Cluster for Pioneering Research, 
\\
Hirosawa, Wako-shi, Saitama 351-0198, Japan}





\begin{abstract}

Hot corino chemistry and warm carbon chain chemistry (WCCC) are driven by gas-grain interactions in star-forming cores: radical-radical recombination reactions to form complex organic molecules (COMs) in the ice mantle, sublimation of CH$_4$ and COMs, and their subsequent gas-phase reactions. These chemical features are expected to depend on the composition of ice mantle which is set in the prestellar phase. We calculated the gas-grain chemical reaction network considering a layered ice-mantle structure in star-forming cores, to investigate how the hot corino chemistry and WCCC depend on the physical condition of the static phase before the onset of gravitational collapse. We found that WCCC becomes more active, if the temperature is lower, or the visual extinction is lower in the static phase, or the static phase is longer.
Dependence of hot corino chemistry on the static-phase condition is more complex. 
While CH$_3$OH is less abundant in the models with warmer static phase, some COMs are formed efficiently in those warm models, since there are various formation paths of COMs.
If the visual extinction is lower, photolysis makes COMs less abundant in the static phase. Once the collapse starts and visual extinction increases, however, COMs can be formed efficiently. Duration of the static phase does not largely affect COM abundances. Chemical diversity between prototypical hot corinos and hybrid sources, in which both COMs and carbon chains are reasonably abundant, can be explained by the variation of prestellar conditions.
Deficiency of gaseous COMs in prototypical WCCC sources is, however, hard to reproduce within our models.

\end{abstract}



\section{Introduction} \label{sec:intro}

Various molecular emission lines have been detected in the central regions of low-mass protostellar cores.
Emission lines of complex organic molecules (COMs), which are defined as organic molecules with 6 atoms or more, are detected in the hot ($\gtrsim 100$ K) central ($\lesssim 100$ au) region in several protostellar cores, e.g.  IRAS 16293-2422B and NGC 1333-IRAS4A \citep[e.g.][]{vandishoeck95, cazaux03, bottinelli04a, bottinelli04b, ceccarelli07, caux11, taquet15, jorgensen16,oya16, lopez17}.
Towards IRAS 04368+2557 in L1527 and IRAS 15398-3359, on the other hand, unsaturated carbon chains (e.g. C$_2$H and C$_4$H) are abundantly detected in the vicinity of protostar \citep{sakai08, sakai09a}. In this work we reserve the term carbon chains for such significantly unsaturated carbon chains to discriminate them from hydrocarbons, which are molecules made of C and H in general.

In theoretical models, the formation of both carbon chains and COMs is explained, at least qualitatively, by the gas-grain chemistry in the sequence of star formation.
In cold ($\sim 10$ K) prestellar cores, molecules with heavy-element such as CO are frozen onto grains. Hydrogen atoms can thermally migrate and hydrogenate CO and other adsorbed atoms and molecules on grain surfaces. While the successive hydrogenation of CO produces
CH$_3$OH, the intermediate radicals such as HCO can also react with each other, if they are in neighboring adsorption sites, to form more complex organic molecules. Eventually, star-formation starts and dust temperature gradually rises, which enhances thermal diffusion of adsorbed species and reactions among them to produce various COMs. Icy molecules are sublimated according to their volatility, and reactions proceed in the gas-phase as well. When the dust temperature exceeds $\sim 100$ K, the dominant constituent of ice, H$_2$O, sublimates together with COMs \citep{rodgers03,garrod06, aikawa08,herbst09,aikawa12, garrodCR13,taquet14,chuang16,cuppen18,lu18}. The low-mass protostars with bright emission of COMs at the central hot region ($R \lesssim 100$) are called hot corinos.
The unsaturated carbon chains, on the other hand, are formed by the gas-phase reactions triggered by the sublimation of CH$_4$ at $\sim 25-30$ K; it is called warm carbon chains chemistry (WCCC) \citep{sakai08,aikawa08,hassel08,sakai13}. Protostars are called WCCC source, if the emission lines of unsaturated carbon chains become bright inward of $\sim$ a few $10^3$ au suggesting their abundance jump.

In theory, both WCCC and COM formation proceed in a star forming core. Indeed, both carbon chains and COMs are recently detected in B335 and L483 \citep{imai16,oya17}, which are called {\it hybrid} (of hot corino and WCCC). The emission of carbon chains is more extended ($r \sim$ a few 100 $- 10^3$ au ) than that of COMs ($r \lesssim 100$ au), which supports the theoretical model. It should be noted, however, that the prototypical hot corino sources are deficient in carbon chains, while COMs are
deficient in the first two WCCC sources.
The column density ratio of C$_2$H/CH$_3$OH is $\sim 0.1$ in prototypical hot corinos such as IRAS 16293 and NGC 1333 IRAS 4A, while it is $\sim 4$ in L1527 \citep[][and references therein]{higuchi18,bouvier20}. 
 \cite{higuchi18} observed multiple lines of C$_2$H, c-C$_3$H$_2$ and CH$_3$OH towards 36 Class 0/I protostars in Perseus molecular cloud complex using IRAM 30 m and NRO 45 m telescopes to derive the chemical composition averaged over the 1,000 au scale. The column density ratio of C$_2$H/CH$_3$OH varies by two orders of magnitude among sources. 
One caution in such single dish observations is a possible contamination of the emission of a parental molecular cloud \citep{bouvier20}.
Follow-up observations are necessary to disentangle the molecular cloud and protostellar core and to confirm the abundance jump of COMs and unsaturated carbon chains, which
characterizes the hot corino and WCCC.  In the ALMA observation of IRAS 16293 B, for example, C$_2$H emission is detected, but its spatial distribution is different from that of C$_3$H$_2$ \citep{murillo18}. It is indeed unexpected in WCCC, and thus suggests the chemistry in IRAS 162293 B would be different from that of prototypical WCCC.

What is the origin of such chemical variation among protostellar cores?
Since the sublimation temperature of COMs are higher than that of CH$_4$, WCCC cores could be deficient in COMs, if the central protostar is not bright enough. But it cannot explain the deficiency of carbon chains (or lack of WCCC) around hot corinos.
\cite{higuchi18} showed that the column density ratio of C$_2$H/CH$_3$OH in protostellar cores does not correlate with the evolutionary indicator, such as bolometric temperature, or with luminosity.
Alternatively, the chemical composition of protostellar cores could reflect the ice composition set in the prestellar phase. \cite{graninger16} observed C$_4$H and CH$_3$OH towards 16 deeply embedded low-mass protostars using IRAM 30 m telescope, and found that the gaseous C$_4$H/CH$_3$OH abundance ratio tentatively correlates with the CH$_4$/CH$_3$OH ice abundance ratio determined by Spitzer c2d survey \citep{boogert08,oberg08}.
The ice composition, in turn, would be determined by the physical conditions. For example, if a core is located near the periphery of a molecular cloud, the penetrating UV radiation keeps carbon atoms abundant, which activates the formation of hydrocarbons.
Then the ratio of CH$_4$/CH$_3$OH in the ice mantle could be relatively high compared with that in cores in the high visual extinction. While the correlation of ice abundances with UV radiation is hard to investigate observationally, the correlation between gaseous molecules with UV radiation field is found in low-mass star-forming cores. \cite{lindberg15} showed that in the protostellar core R CrA IRS7B, COMs are under-abundant, while CN emission is strong, which is an indicator of PDR chemistry. \cite{higuchi18} found that the protostars located near cloud edges or in isolated clouds tend to have a high C$_2$H/CH$_3$OH ratio.
In a prototypical prestellar core, L1544, \cite{spezzano16} found that c-C$_3$H$_2$ emission peaks close to the southern part of the core, where the surrounding molecular cloud has a sharp edge, while CH$_3$OH mainly traces the northern part of the core. The gas in north and south would eventually fall to the central region to set the chemistry in protostellar phase.
Since a protoplanetary disk is being formed in the central region of a protostellar core, the chemical composition of a core might be inherited by a forming disk and eventually by a planetary system.

Several theoretical work investigated the dependence of molecular abundances on core models. \cite{garrod06} showed that grain-surface reactions become more important for COM formation in models with slower warming-up, while the peak abundance of each molecular species show complex dependence on the warm-up timescale. \cite{aikawa12} showed that for WCCC to be active, the gas density around the CH$_4$ sublimation zone should be low enough for C$^+$ to be abundant. \cite{sakai13} pointed out that CH$_4$ should be more abundant than $10^{-7}$ relative to hydrogen to be the major reactant with C$^+$ (competing against C$^+$ + OH) to trigger WCCC. But there are few theoretical work which investigates both WCCC and hot corino chemistry systematically. In this paper, we study the dependence of COMs and warm carbon-chain abundances in protostellar cores on the physical conditions in prestellar phase via numerical calculations. Specifically, we vary the temperature, visual extinction of the ambient clouds, and duration of static phase before gravitational collapse. 

The rest of the paper is organized as follows. Our chemical and physical models are described in \S 2. Results of model calculations are presented in \S 3. We compare our results with previous work and observations in \S 4. Our conclusions are summarized in \S 5.

\section{Model}

\subsection{Model of a Star-forming Core}

Structure and evolution of molecular clouds are  summarized as follows \citep[][and references therein]{andre14}.
Molecular clouds consist of filaments. Herschel observations revealed that gravitationally bound prestellar cores and protostars are primarily found in filaments with the H$_2$ column densities of $\gtrsim 7\times 10^{21}$ cm$^{-2}$.
The radial density profile of the filaments show a flat plateau with FWHM $\sim 0.1$ pc. The mean gas density in the star-forming filaments is thus $n$(H$_2$)$\gtrsim 2\times 10^4$ cm$^{-3}$. Statistical observations of cloud cores suggest the lifetime of a core with $n$(H$_2$)$\sim 10^4$ cm$^{-3}$ to be $10^6$ yr, which coincides with the sound-crossing time scale of a filament. 
The typical column density of filaments with cores roughly coincides with the threshold value of gravitational instability, which indicates that the filaments fragment to form self-gravitating cores. Eventually the core collapses to form protostars.

\begin{figure}[ht!]
\plotone{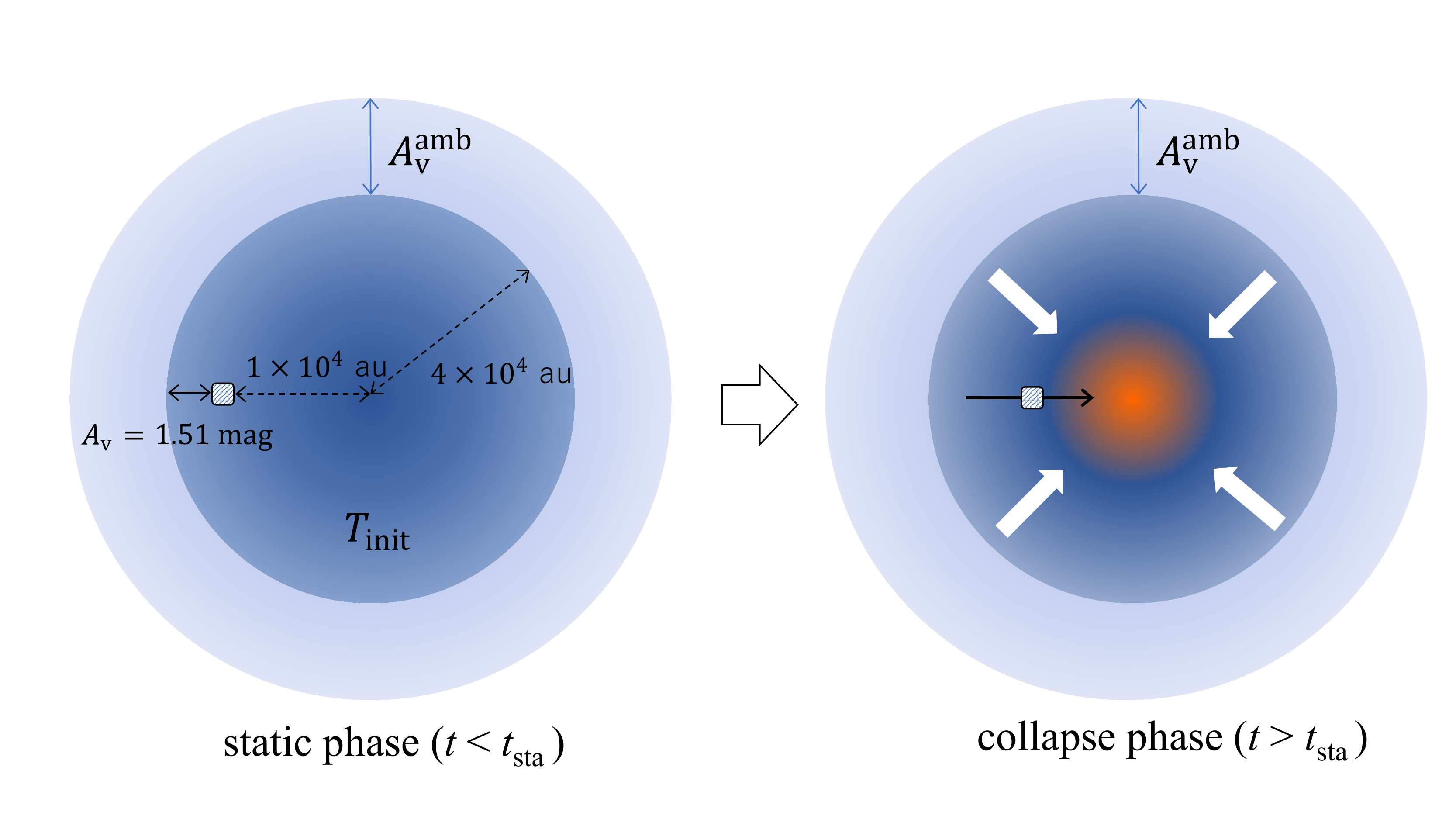}
\caption{Schematic view of our core model (see text). The radial size is not to scale.
\label{fig_schem}}
\end{figure}

In the present work, we solve the rate equations of gas-grain chemical reaction network in a fluid parcel  that reaches the $T\gtrsim 100$ K region in a protostellar envelope.
Considering the observational overview described above, we divide our model to two phases: the prestellar phase with density of $\sim 10^4$ cm$^{-3}$ and the collapse phase (Figure \ref{fig_schem}).

For the latter phase, we adopt 
the 1D (spherical) radiation hydrodynamic model of low-mass star formation by \cite{masunaga98} and \cite{masunaga00}. We assume the spherical collapse,
although flattened structure will form in the vicinity of the protostars \citep[e.g.][]{terebey84}.
The total mass and initial radius of the core are 3.852 $M_{\odot}$ and $R=4\times 10^4$ au, respectively, The initial number density of hydrogen nuclei is $\sim 6\times 10^4$ cm$^{-3}$ (i.e. $n$(H$_2$)$\sim 3\times 10^4$ cm$^{-3}$) at the core center.
After the onset of gravitational collapse, the central density increases with time, and the protostar is formed in $2.5\times 10^5$ yr. After the birth of the protostar,
the model further follows the evolution for $9.3 \times 10^4$ yr, during which the protostar grows by mass accretion from the envelope. At each evolutionary stage, the model gives the total luminosity of the core and the radial distribution of density, temperature, and infall velocity self-consistently. At the final time step, the total luminosity of the core is 24 $L_{\odot}$ and the temperature is higher than 100 K inside $R \sim 100$ au.  While the temperatures of gas and multiple components of dust material are calculated separately in the original model of \cite{masunaga98} and \cite{masunaga00}, we assume that the dust temperature is equal to the gas temperature, for simplicity. More detailed description of our core model can be found in \cite{masunaga98, masunaga00, aikawa08, aikawa12}.

\begin{figure}[ht!]
\plotone{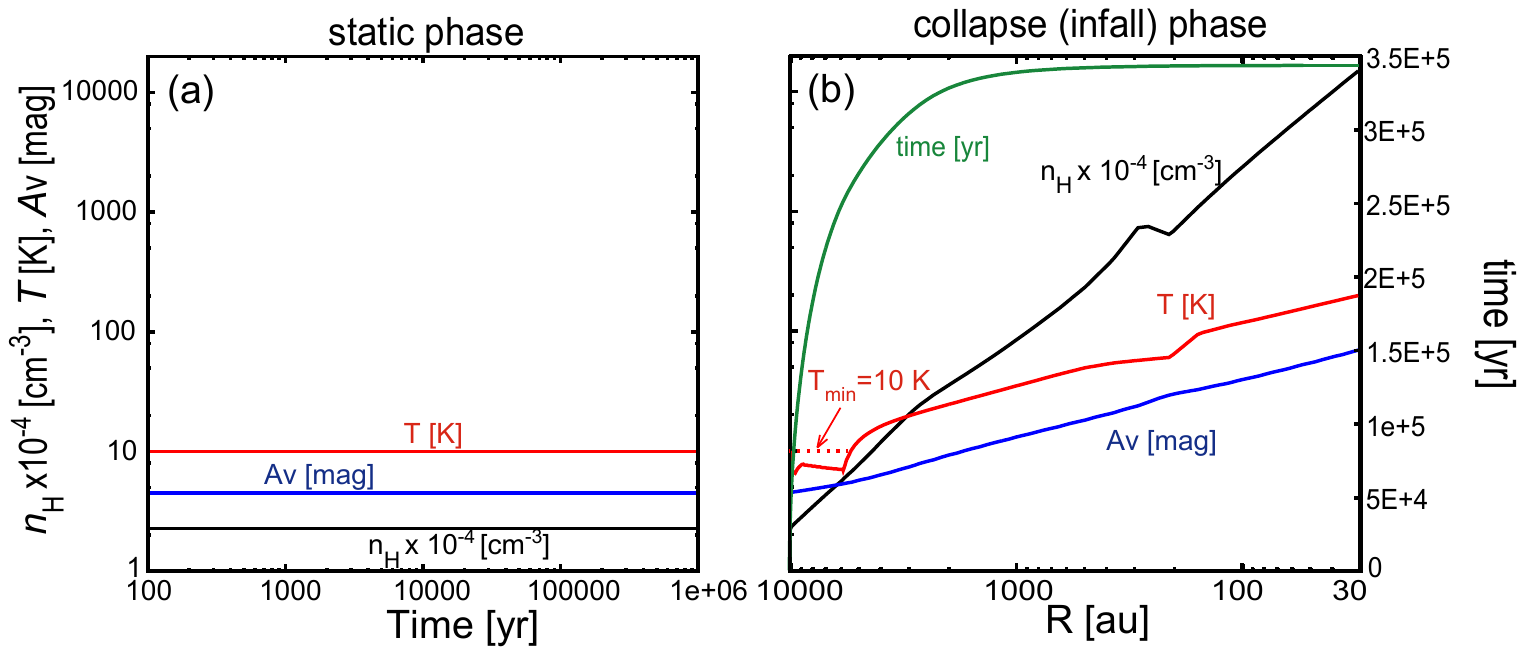}
\caption{
(a) The physical parameters (gas number density in black, temperature in red, and visual extinction in blue) in the static phase of our fiducial model. While they are plotted as a function of time, the values are constant in the static phase.
(b) The physical parameters in the infalling fluid parcel in the collapse phase as a function of the radial location of the fluid parcel (see text) in our fiducial model. The time after the onset of collapse is depicted by the green line. The red dotted line depicts the temperature in the model with $T_{\rm min}=10$ K.
\label{phys_fid}}
\end{figure}

In our previous work \citep{aikawa08, aikawa12}, we solved the rate equations in multiple infalling fluid parcels to derive the radial distributions of molecular abundances.
In the present work, on the other hand, we follow the temporal variation of molecular abundances in one infalling fluid parcel that is initially located at $R=1.01\times 10^4$ au and reaches $R=30.6$ au at the final time step, and plot the abundances as a function of the temporal radial position of the fluid parcel.
The green line in Figure \ref{phys_fid} (b) shows the time after the onset of collapse as a function of the location of the infalling fluid parcel. The figure also depicts the density, temperature, and visual extinction of the fluid parcel.
As we will show in section \S 3, the $R-$abundance plot is reasonably similar to the true radial distribution of molecules around WCCC and hot corino regions at the final time step. If we were to calculate the chemistry in multiple fluid parcels to derive the radial distributions with a similar spatial resolution as the $R-$ abundance plot, we would need $\sim 100$ fluid parcels for each model, which is computationally expensive. We decided to use the $R-$abundance plot as a proxy of the radial distribution, considering the similarity between them.

For the prestellar phase, we assume that the initial core stays static for $t_{\rm sta}=1\times 10^6$ yr.
We thus call this phase as static phase in the following. This notation, static phase versus collapse phase, is clearer, since a protostar is not formed immediately after the onset of collapse; i.e. the early collapse phase is actually a prestellar phase.
In the fluid parcel we follow, which is located at  $R=1.01\times 10^4$ au, the density and temperature are $n_{\rm H}=2.28\times 10^4$ cm$^{-3}$ and $T_{\rm init}=10$ K  (Figure \ref{fig_schem} and Figure \ref{phys_fid} a). 
The gas column density from the core outer boundary ($R=4\times 10^4$ au) to the location of the fluid parcel corresponds to the visual extinction of $A_{\rm v}=1.51$ mag. Outside the core, we assume the ambient gas of $A_{\rm v}^{\rm amb}=3$ mag. The interstellar UV radiation field is thus attenuated by the total visual extinction of 4.51 mag (Figure \ref{phys_fid} a).
These parameters are chosen to be consistent with the physical conditions in filaments and for a simple and smooth transition to the infalling stage.
Ideally, we need to also consider the formation of a filament and its fragmentation to form prestellar cores. We will include the filament formation to our model in \S 3.5.


In order to investigate the dependence of WCCC and COM formation on the physical parameters of static phase, we vary the initial temperature ($T_{\rm init}=10, 15, 20,$ or 25 K), the visual extinction of the ambient gas ($A_{\rm v}^{\rm amb}=1, 3,$ or 5 mag) and duration of the static core phase ($t_{\rm sta}=3\times 10^5, 1\times 10^6,$ or $3\times 10^6$ yr) (Table \ref{tab:param}). In the collapse phase of these models,  the temporal variation of the temperature is assumed to be the same as the fiducial model, while the visual extinction is modified in accordance with $A_{\rm v}^{\rm amb}$. Namely, even in the model with high $T_{\rm init}$, the temperature of the fluid parcel decreases below 10 K in the early phase of collapse, shielded from the interstellar radiation field.  As we will see in \S 3, the ice composition is significantly changed during this period, which dilutes the effect of warm static phase in the models of $T_{\rm init}>10$ K.
In order to further investigate the effect of thermal history, we also calculated models in which the minimum temperature is set to be $T_{\rm min}=10, 15, 20$ or 25 K;
the temperature of the fluid parcel is fixed at $T_{\min}$ in the static phase and early collapse phase, and start rising when it exceeds $T_{\rm min}$ in the original model.  The dotted line in Figure \ref{phys_fid} (b), for example, depicts the temporal temperature variation in the model of $T_{\rm min}=10$ K. Warm temperatures in the static and early collapse phase would be possible, if the core is in e.g., cluster-forming regions.

\begin{deluxetable}{llc}
\tablecaption{Physical Parameters of the Static Phase \label{tab:param}}
\tablehead{\colhead{fixed parameters} &\colhead{} & \colhead{}  }
\startdata
total mass of the core & & 3.852 [$M_{\odot}$]\\
core radius & & $4\times 10^4$ [au] \\
initial position of the infalling fluid parcel & & $1.01 \times 10^4$ [au]\\
final position of the infalling fluid parcel & & 30.6 [au] \\
\hline
varied parameters\tablenotemark{a}: & & \\
temperature in the static phase & $T_{\rm init}$ & {\bf 10}, 15, 20, 25 [K] \\
minimum temperature & $T_{\rm min}$ & 10, 15, 20, 25 [K] \\
visual extinction of the ambient gas & $A_{\rm v}^{\rm amb}$ & 1, {\bf 3}, 5 [mag]\\
duration of static phase & $t_{\rm sta}$ & $3\times 10^5$, {\boldmath $1\times 10^6$}, $3\times 10^6$ [yr]\\ 
\enddata
\tablenotetext{a}{The fiducial values are in bold font}
\end{deluxetable}

\subsection{Chemistry}
Our reaction network model is based on \cite{garrod13} and includes the following updates. The gas-phase reaction of H$_2$CO + OH has two production paths: HCOOH + H and H$_2$O + HCO. In our previous model \citep{aikawa12}, the former was one of the major formation paths of HCOOH in the gas phase. According to quantum calculations and laboratory experiments \citep{,d'anna03,sivakumaran03,xu06}, however, the former branch has the activation barrier of  $\sim 2500$ K and thus negligible compared with the latter branch, which is barrierless. We thus deleted the former path in the present work, while the rate coefficient of the latter path is set to be $1.0\times 10^{-11}(T/300.0 {\rm K)}^{-0.6}$ cm$^3$ s$^{-1}$ \citep[see also][]{ruaud15}.
While HCOOH consists of 5 atoms, we consider it as COM in the present work, since it is often observed in hot corinos \citep[e.g.][]{remijan06, imai16}.
The gas-phase reaction rate coefficient  of NH$_2$ + H$_2$CO $\rightarrow$ NH$_2$CHO + H is set to be $2.6 \times 10^{-12} (T/300.0 {\rm K})^{-2.1} \exp(-26.9 {\rm K}/ T)$ cm$^3$ s$^{-1}$ following \cite{barone15}. On grain surfaces, the reaction of H + NH$_2$CO results in hydrogen abstraction (H$_2$ + HNCO) rather than hydrogenation (NH$_2$CHO) \citep{noble15}.
The grain-surface reaction of HCO + CH$_3$ has three possible production paths, CH$_4$ + CO, CH$_3$CHO, and CH$_3$OCH. We neglected the latter two paths following \cite{enrique-romero16}, who showed that only the first path is plausible considering the orientation of the reactants on amorphous water ice (see \S 4.2 for further discussions).
We also added the gas-phase reactions of H$_2$ with C$_2$H, C$_4$H and C$_3$H$_2$, which result in a hydrogen addition to the carbon chains (i.e. C$_x$H$_y$ + H$_2 \rightarrow$ C$_x$H$_{y+1}$ + H ); the barrier is set to be 998 K, 950 K, and 1740 K, respectively, referring the KIDA data base for astrochemistry (kida.obs.u-bordeaux1.fr).

The simplest model of gas-grain chemistry consists of two phases, gas phase and ice phase, and thus is called two-phase model \citep{hase92}. While the ice mantle consists of $\sim 10^2$ monolayers of molecules and atoms in cold dense prestellar cores, such layering structure is not considered in the two-phase model. For example, all CO molecules in the ice mantle have the equal probability of reaction and the equal probability of desorption. It is obviously a simplification. \cite{hase93} then proposed the three-phase model that consists of gas-phase, ice surface phase, and ice mantle phase. In the original three-phase model, the ice mantle phase is assumed to be chemically inert, and chemical reactions are considered only in the ice surface phase. In more recent three-phase models, the chemical reactions are often considered in the ice mantle phase as well, but with slower rates than in the ice surface phase. The basic assumption of the three-phase model is that the ice mantle phase has a uniform chemical composition, which is not true. Infrared absorption bands of interstellar ices suggest that ice mantle is made of polar (water-rich) component and apolar (CO- and CO$_2$-rich) component \citep[e.g.][]{gibb04}. In order to take into account such inhomogeneity in the ice mantle, we adopt a seven-phase model, that consists of the gas phase, the ice surface phase, and 5 phases of ice mantle. We assume that the top 4 monolayers are the ice surface phase \citep{vasyunin13}, while each mantle phase consists of a few tens of monolayers at a maximum.   Since the term {\it phase} is used to discriminate gas and ice, i.e. the gas phase versus ice phase, and also to specify the evolutionary stage of a star-forming core, i.e. the static phase and infalling phase, in the present work, we use the term {\it layer} to specify different phases of ice hereinafter. In summary, ices in our model consists of six layers: the ice surface layer and 5 mantle layers.

We assume the Langmuir-Hinshelwood mechanism for reactions in the ice surface layer and mantle layers; species can diffuse by thermal hopping and react with each other when they meet. No quantum tunneling is assumed in the migration, even for H atoms, while we consider the tunneling effect for reaction probability of reactions with activation barrier.
The barrier of thermal migration of atoms and molecules are set to be 40 \% of the adsorption energy $E_{\rm ads}$ in the surface layer, while it is 80 \% of $E_{\rm ads}$ in the ice mantle layers \citep[e.g.][]{ruaud16}. The set of adsorption energies is adopted from \cite{garrod13}. 
We consider two-body reactions in each ice layer, but do not allow the reactions between species in different layers, assuming the vertical migration of species in ice mantles is limited.
Swapping between the six ice layers is not considered. Only the molecules in the surface layer are subject to desorption, while the species in the deeper mantle layers are transported to the upper layers following the net loss of ices in the surface layer. Detailed explanation on the formulation and coding is given in \cite{furuya17}.
The molecules in the ice mantle are subject to photolysis with extinction in the pre/protostellar core and the outer layers of ice mantles.

As elemental abundances, we adopt the so-called low-metal values \citep[Table 1 of][]{aikawa01}.
In our fiducial model, the species are assumed to be initially in the form of atoms or atomic ions except for hydrogen, which is entirely in its molecular form, in the static core. 
Later in section 3, we also calculate models in which the initial molecular abundance of the static phase is obtained via the 1-D shock model of molecular cloud (i.e. filament) formation \citep{bergin04, furuya15}. 

\section{Result}
\subsection{Fiducial Model}

\begin{figure}[ht!]
\includegraphics[scale=0.85]{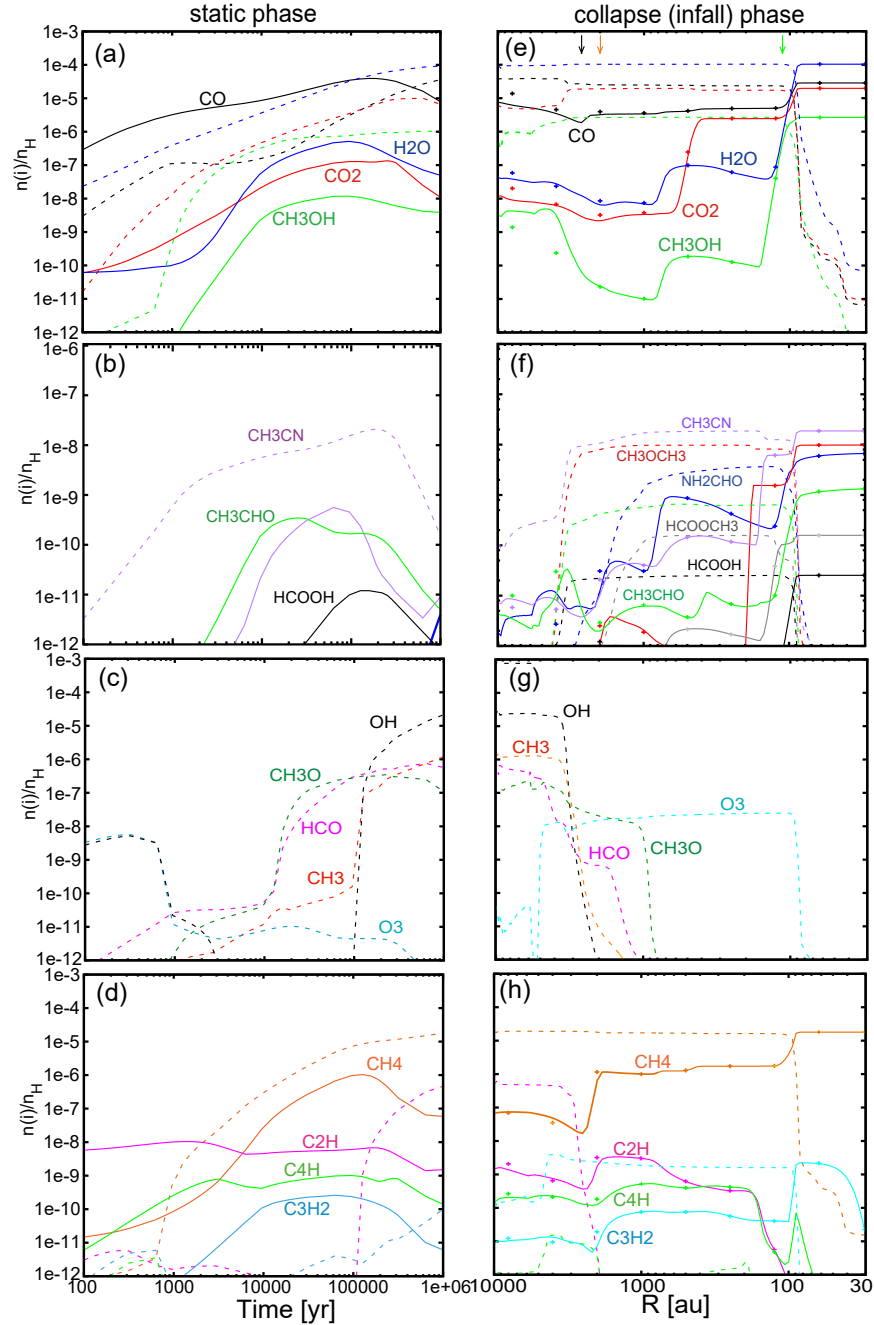}
\caption{
{\it Left column}: The molecular abundances relative to hydrogen nuclei as a function of time in the static phase in our fiducial model. The solid lines show the gas-phase abundances, while the dotted lines depict the ice abundances. 
{\it Right column}: The molecular abundances in the infalling fluid parcel in the collapse phase as a function of the radial location of the fluid parcel (see text) in our fiducial model.
The arrows in panel (e) depict the radius where the thermal desorption timescale is equal to adsorption timescale for CO (black), CH$_4$ (orange) and CH$_3$OH (green).
The crosses in panel show the radial distributions of gaseous molecules at the final time step in our model.
\label{dist_fid}}
\end{figure}

Figure \ref{dist_fid} (a)-(d) shows the temporal variation of molecular abundances in the static prestellar core phase. The solid lines depict the molecular abundances of gaseous molecules relative to hydrogen nuclei, while the dashed lines show those of icy molecules, i.e. the sum of ices in the surface layer and mantle layers. As we start with atoms and ions, C$^+$ is gradually converted to atomic carbon and then to CO in the gas phase, while oxygen is hydrogenated to form water on grain surfaces (Figure \ref{dist_fid} a). After $10^5$ yrs, the ice abundances of CO, CO$_2$, and CH$_3$OH relative to water are in reasonable agreement with observation towards low-mass protostars and background stars (i.e. CO/H$_2$O$\sim 0.3$, CO$_2$/H$_2$O $\sim 0.3-0.4$, and CH$_3$OH/H$_2$O $\sim$ a few $10^{-2}$) \citep{oberg11}. Hydrocarbons, including carbon chains, are efficiently formed from C$^+$ and atomic C both in the gas phase and ice before carbon is fully converted to CO (Figure \ref{dist_fid} d). Hydrocarbons and carbon chains also react with N atoms and O atoms to form CH$_3$CN and CH$_3$CHO in the gas phase (Figure \ref{dist_fid} b). CH$_3$CHO, for example, is mainly formed via O + C$_2$H$_5$ at $t\sim 10^4$ yr in our model. 

Molecular evolution in the collapse phase is shown as a function of the radial position of the infalling fluid parcel in Figure \ref{dist_fid} (e)-(h). The sublimation temperatures of the most volatile C-bearing molecules, CO and CH$_4$, are $\sim 20-30$ K, depending of gas density, which is reached when the fluid parcel is at $R=2000-3000$ au. Since the ice mantle is multi-layered and made of mixture of ices in our model, only a fraction of CO and CH$_4$ is sublimated. The majority of icy molecules are trapped in water ice until water is sublimated at $\sim 100$ K. This entrapment is in agreement with laboratory experiments, at least qualitatively, while the actual fraction of trapped volatiles depends on the ice thickness and the mixture ratio in experiments, and parametrization of diffusion and swapping efficiency in numerical simulations \citep{collings04, fayolle11}.

Figure \ref{dist_fid} (c) and (g) show the abundances of several icy radical species in the static phase and collapse phase, respectively. Radicals are efficiently formed, trapped, and stored in ice in the cold phase \citep{lu18}.
Compared with the two-phase model, various radicals are abundant in the multi-layered ice mantle model for the following reasons.
Firstly, icy species, including radicals, are trapped beneath the surface layer, and not directly subject to desorption to the gas phase. Radicals beneath the surface layer do not react with newly adsorbed atoms and molecules from the gas phase, either. In the two-phase model, on the other hand, all icy radicals can react with newly adsorbed species, e,g. H atoms, to form saturated molecules. Lastly, the thermal diffusion and thus the reactions (e.g. radical-radical reactions) are slower in ice mantle layers than in the surface layer. When the temperature rises, however, the stored radicals start to react with each other to form COMs. 
CH$_3$OCH$_3$, for example, is mainly formed in the ice mantle via the reaction of CH$_3$ + CH$_3$O, when the temperature of the fluid parcel is $\sim 20$ K (Figure \ref{dist_fid} f). The CH$_3$OCH$_3$ abundance in the final time step in the present model ($9.7 \times 10^{-9}$) is much higher than that in the two-phase model of \cite{aikawa12} ($1.2\times 10^{-10}$) adopting the same collapse model.   
The gas-phase reactions of sublimated molecules also contribute to the formation of COMs, e.g. C$_2$H$_5$ + O $\rightarrow$ CH$_3$CHO + H and NH$_2$ + H$_2$CO $\rightarrow$ NH$_2$CHO + H at $R\sim 90$ au.

When the fluid parcel reaches $R\sim 2000$ au, the temperature rises to $\sim 25$ K, which is the sublimation temperature of CH$_4$. Sublimation of CH$_4$ triggers the WCCC; e.g. CH$_4$ reacts with C$^+$ to produce C$_2$H$_3^+$, which then recombine to form C$_2$H. The carbon chain also extends via the reactions with C atoms (e.g. C + CH$_3$ $\rightarrow$ C$_2$H$_2$ + H).
It should be noted that CH$_4$ does not fully sublimate at its own sublimation temperature ($\sim 25$ K), which suppresses the WCCC compared with the two-phase model of \cite{aikawa08}.
Closely looking at Figure \ref{dist_fid} (f) and (h), we also note that WCCC enhances the abundance of some COMs, as well. For example, the NH$_2$CHO abundance  increases around the WCCC region ($R\sim 1900$ au); CH$_3$ reacts with O atoms in the gas phase to form H$_2$CO, which then reacts with NH$_2$ to form NH$_2$CHO. Later, the temperature rises to the sublimation temperature of H$_2$CO ($\sim 40$ K) when the fluid parcel reaches $\sim 750$ au. Then the gas-phase reactions initiated by the sublimation of H$_2$CO further enhance the abundance of NH$_2$CHO.

The crosses in Figure \ref{dist_fid} (e), (f), and (h) depict the radial distribution of gaseous molecular abundances at the final time step; we calculated the gas-grain chemistry in the fluid parcels that reach the radius of 62.4 au, 125 au, 250 au, 500 au, 1000 au, 2000 au, 4000 au and 8000 au, respectively.  
The $R$-abundance plot of the infalling fluid parcel (i.e. solid lines) is quite similar to this true radial distribution inside the radius of $\sim 10^3$ au. It is reasonable, because the infall timescale inside $10^3$ au is $\lesssim 5\times 10^3$ yr, which is much shorter than the age of the protostar $9\times 10^4$ yr in our model. At $R\gtrsim 10^3$ au, the difference between the $R-$abundance plot and the true radial distribution becomes more apparent. In the WCCC region ($r\sim$ a few $10^3$ au),  however, the $R-$abundance plot of CH$_4$ and hydrocarbons are in reasonable agreement with the radial distribution. Thus we can use the $R-$abundance plot to investigate the hot corino chemistry and WCCC.

\subsection{Dependence of COM Abundances on $T_{\rm init}$ and $T_{\rm min}$}

\begin{figure}[ht!]
\plotone{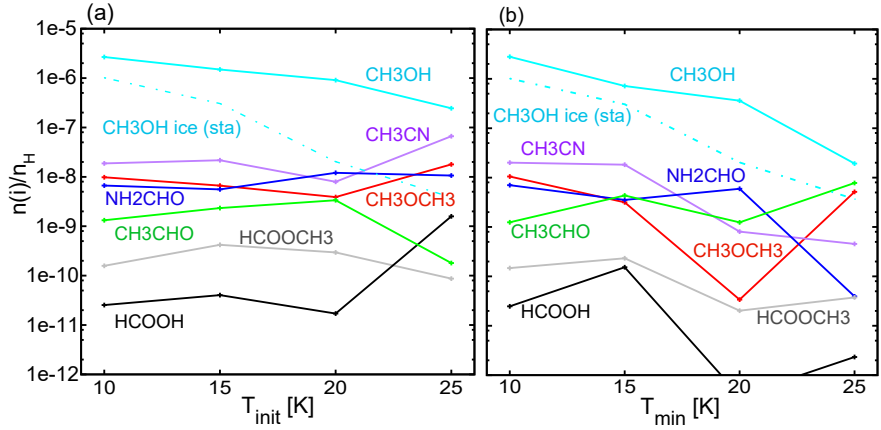}
\caption{Gas phase abundances of COMs at the final time step ($R=30.6$ au) in models with the initial temperature of 10 K, 15 K, 20 K, and 25 K (a),
and in models with the minimum temperature of 10 K, 15 K, 20 K, and 25 K (b). The dash-dotted cyan lines depict the CH$_3$OH ice abundance at the end of the static phase.
\label{Tinit}}
\end{figure}

We calculate the molecular evolution in the fluid parcels with the temperature in the static phase $T_{\rm init}$  of 15 K, 20 K, and 25 K. The gaseous abundances of COMs in the final time step
(i.e. $R=30.6$ au) are shown in Figure \ref{Tinit} (a) as a function of $T_{\rm init}$. A naive expectation is that the higher $T_{\rm init}$ makes the freeze out and grain-surface hydrogenation less efficient, resulting in lower abundances of icy molecules and COMs. Indeed, the abundances of CH$_3$OH and CH$_3$CHO are lower in the model with $T_{\rm init}=25$ K than in the fiducial model (i.e. $T_{\rm init}=10$ K). But the abundances of some COMs, such as CH$_3$CN, CH$_3$OCH$_3$ and HCOOH, are higher with $T_{\rm init}=25$ K than in the fiducial model.

\begin{figure}[ht!]
\plotone{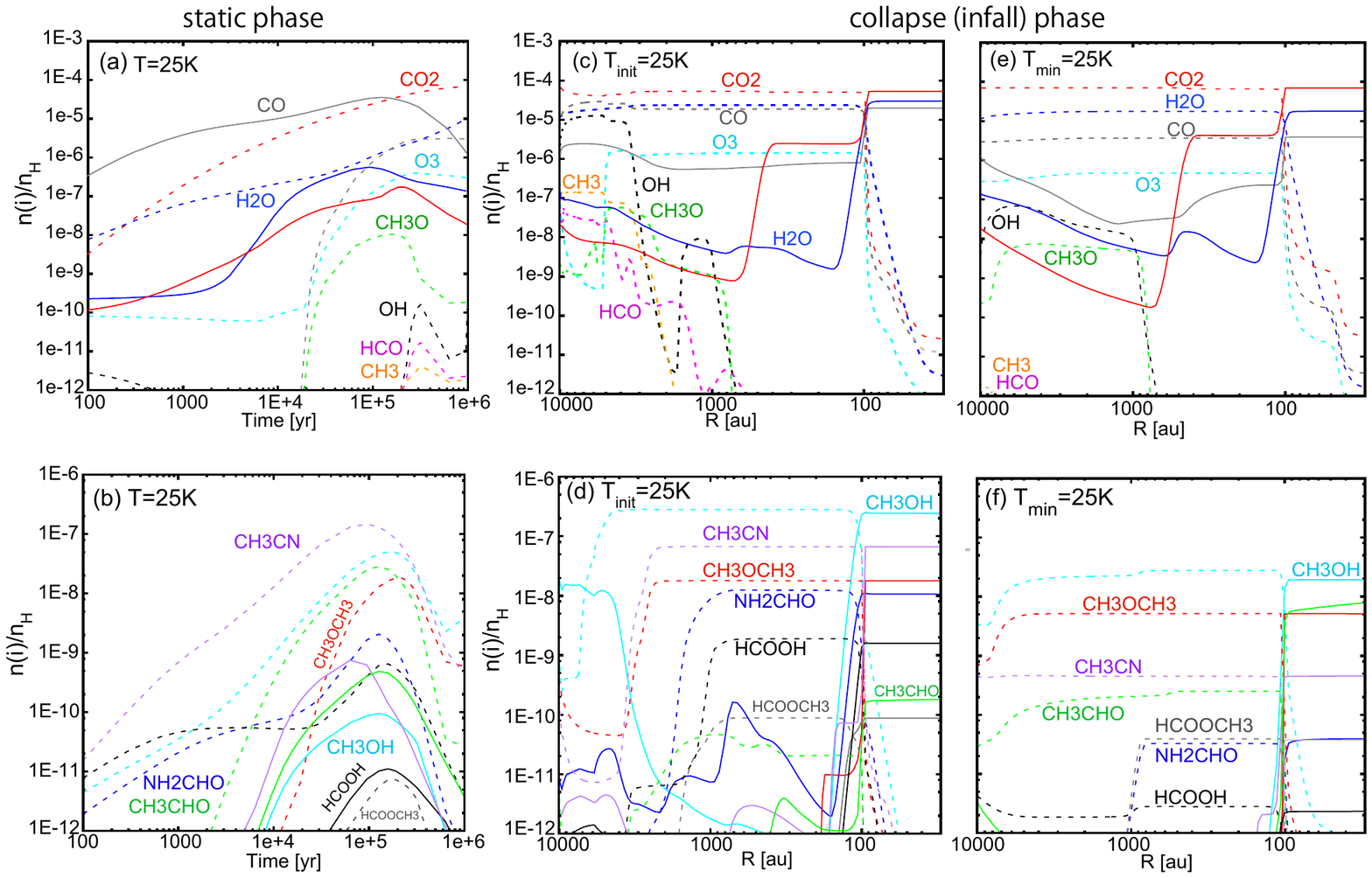}
\caption{(a, b) Temporal variation of abundances of gaseous (solid lines) and icy (dashed lines) radicals and COMs relative to hydrogen nuclei in the static phase with $T_{\rm init}=25$ K. The abundances of CO and H$_2$O are also plotted as a reference.
(c, d) Abundances of radicals and COMs in the infalling fluid parcel as a function of the radial location of the fluid parcel in the model with $T_{\rm init}=25$ K.
(e, f) Abundances of radicals and COMs in the infalling fluid parcel as a function of the radial location of the fluid parcel in the model with $T_{\rm min}=25$ K.
\label{warm_COMs}}
\end{figure}

In order to investigate the chemistry with high $T_{\rm init}$, we plot the temporal variation of molecular abundances in the model with $T_{\rm init}=25$ K;  Figure \ref{warm_COMs} (a, b) shows  the temporal variation in the static phase, while Figure \ref{warm_COMs}(c, d) is the $R-$abundance plot in the infalling phase. 
In the static phase, the abundances of H$_2$O ice and CO ice are indeed lower than in our fiducial model; at the end of static phase ($t=10^6$ yr), their abundances are $1\times 10^{-5}$ (H$_2$O) and  $3\times 10^{-6}$ (CO). CO$_2$ ice, on the other hand, is abundantly formed via CO + OH on grain surfaces; e.g. its abundance reaches $\sim 7\times 10^{-5}$ at $t=10^6$ yr. The warm temperature makes the CO freeze-out less efficient, but enhances the thermal diffusion of species with moderately high binding energy.
In numerical models of gas-grain chemistry, molecules tend to be converted to and accumulate as stable icy species whose sublimation temperature is higher than the current temperature \citep[e.g.][]{aikawa97, furuya14}. Icy abundances of COMs in the static phase are also higher than those in our fiducial model. While CH$_3$OH is mainly formed via hydrogenation of CO in our fiducial model ($T_{\rm init}=10$ K), it is formed via the grain-surface association of CH$_3$ + O $\rightarrow$ CH$_3$O and subsequent hydrogenation in the $T_{\rm init}=25$ K model. CH$_3$ and O atoms, in turn, are formed via photodissociation of larger hydrocarbons and CO$_2$, respectively. The grain-surface reaction of CH$_3$ with CH$_3$O forms CH$_3$OCH$_3$.  In other words, formation of COMs starts from carbon chains and hydrocarbons, which are abundantly formed in both the gas and ice phases before carbon is fully converted to CO. Once CO becomes the dominant carbon reservoir
($t\sim 10^5$ yr), COM abundances temporally decline, as the destruction (e.g. photodissociation) dominates over the formation.

The model results are in line with recent laboratory experiments showing that CH$_3$OH can be formed in CH$_4$ ice mixture. \cite{qasim18} performed Temperature Programmed Desorption (TPD) of the mixed ice of CH$_4$, O$_2$ and H atoms to find that CH$_3$OH is produced. The analysis of the experimental data with different ice mixtures indicates that CH$_3$OH is formed via CH$_4$ + OH $\rightarrow$ CH$_3$ + H$_2$O and CH$_3$ + OH $\rightarrow$ CH$_3$OH. The latter reaction is one of the main formation paths of CH$_3$OH in the collapse phase of $T_{\rm init}=25$ K model.
In the static phase of our model with $T_{\rm init}=25$ K, CH$_4$ ice abundance is not high, since $T_{\rm init}$ coincides with its sublimation temperature. CH$_3$ is thus formed from larger hydrocarbons, and then quickly reacts with atoms and radicals in the ice mantle.

Figure \ref{warm_COMs} (c, d) is the $R-$abundance plot in the infalling phase of the model with $T_{\rm init}=25$ K. It should be noted that the temperature of the fluid parcel falls to $\lesssim 10$ K at the onset of collapse and the low temperature $\lesssim 20$ K continues for $\sim 3.1\times 10^5$ yr  (\S 2.1). During this period, CH$_3$OH ice is abundantly formed via CO ice hydrogenation, while the thermal diffusion of species with heavy elements and thus radical-radical reactions are quenched. In Figure \ref{Tinit} (a), we can compare the abundances of gaseous CH$_3$OH at the final timestep (i.e. $R=30.6$ au) (solid cyan line) and CH$_3$OH ice at the end of static phase (dash-dotted cyan line); the former is much higher than the latter at $T_{\rm init}\gtrsim 15$ K mainly due to the CH$_3$OH formation during this temporal cold phase.
Radicals are also abundantly stored in the ice mantle then. It makes the dependence of the final COM abundances on $T_{\rm init}$ less significant than that on $T_{\rm min}$ (see below). Yet, chemical signature of warm $T_{\rm init}$ remains in some layers in the ice mantle. For example, CH$_3$CHO is less abundant in the model with $T_{\rm init}=25$ K, since its precursor CH$_3$ in the ice mantle is less abundant. On the other hand, HCOOH is abundantly formed via OH + HCO in the upper layers of bulk ice mantle at $R\sim 1\times 10^3$ au in the model of $T_{\rm init}=25$ K. CO ice is not abundant in those layers, and HCO is formed via reaction of OH + H$_2$CO, while H$_2$CO is formed via photodissociation or H-abstraction of CH$_3$OH.
At the same radius in the fiducial model, OH reacts with CO, which is relatively abundant in all the ice mantle layers.

While it is reasonable that the core temperature decreases as the gas density increases in the early collapse phase, we also calculated models with a fixed minimum temperature. The model with $T_{\rm min}=25$ K, for example,  is similar to the model with $T_{\rm init}=25$ K, but the temperature is kept at 25 K even in the early collapse phase (i.e. $t=1\times 10^6 - 1.33\times 10^6$ yr). 
Figure \ref{Tinit} (b) shows the abundances of gaseous COMs at the final time step, and Figure \ref{warm_COMs} (e, f) shows the $R-$abundance plot of radicals and COMs in the collapse phase of $T_{\rm min}=25$ K model. Note that their abundances in the static phase are the same as Figure \ref{warm_COMs} (a, b). Although the icy abundances of CH$_3$OH and some radicals increase temporally after the onset of collapse, the increment is smaller than in the model with $T_{\rm init}=25$ K (Figure \ref{warm_COMs} c, d).
COM abundances depend more sensitively on $T_{\rm min}$ than on $T_{\rm init}$; e.g. CH$_3$OCH$_3$, which is formed via CH$_3$O + CH$_3$ in the ice mantle, is least abundant with $T_{\rm min}=20$ K, and becomes more abundant with $T_{\rm min}=25$ K. The reactant, CH$_3$O is mainly formed via photodissociation of CH$_3$OH in the low $T_{\rm min}$ model ($<20 K$), while it is formed from hydrocarbon (CH$_3$ + O) in the model of $T_{\rm min}=25$ K. Either path is not efficient in the $T_{\rm min}=20$ K model.

\subsection{Dependence of WCCC on $T_{\rm init}$ and $T_{\rm min}$}

\begin{figure}[ht!]
\plotone{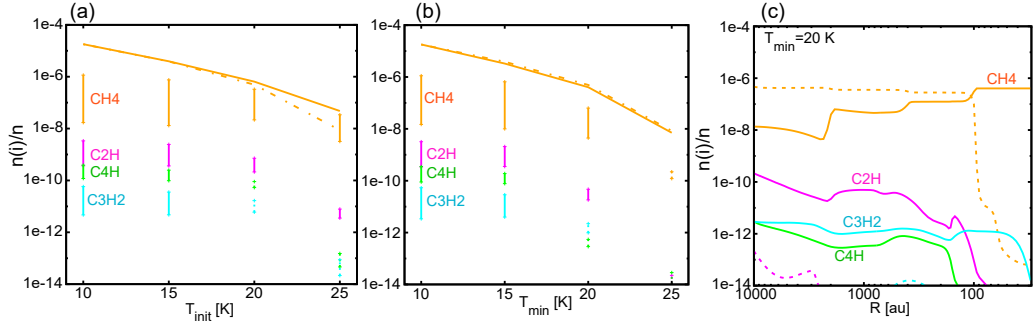}
\caption{(a) The maximum and minimum abundances of gaseous CH$_4$ (orange), C$_2$H (magenta), C$_4$H (green), and C$_3$H$_2$ (cyan) around the CH$_4$ sublimation region ($R=1455-2850$ au) as a function of $T_{\rm init}$. The upper solid orange line depicts the CH$_4$ gas abundance at the final timestep (i.e. $R=30.6$ au), while the dash-dotted orange line depicts the CH$_4$ ice abundance at the end of the static phase. (b) Same as (a) but as a function of $T_{\rm min}$. The maximum and minimum values are connected by solid lines, if the abundance increases inwards (i.e. WCCC), while the abundance decreases inwards for dotted lines.
(c) $R-$abundance plot of CH$_4$ and carbon chains in the collapse phase as in Figure \ref{dist_fid} (h) but in the model of $T_{\rm min}=20$ K. 
\label{Tinit_WCCC}}
\end{figure}

In Figure \ref{Tinit_WCCC} (a)(b), we plot the minimum and maximum abundances of CH$_4$, C$_2$H, C$_3$H$_2$, and C$_4$H around the CH$_4$ sublimation region, which we set $R=1455-2850$ au, in models with $T_{\rm init}=10, 15, 20,$ and 25 K (panel a) and $T_{\rm min}=10, 15, 20,$ and 25 K (panel b). The minimum and maximum values are connected by a vertical solid line, if the abundance increases with decreasing radius in the CH$_4$ sublimation region, which indicates that WCCC is active. A dotted vertical line, on the other hand, is used if the abundance decreases inwards. We can see that CH$_4$ abundance decreases with increasing $T_{\rm init}$ and $T_{\rm min}$; it is natural, since both the grain surface formation and freeze-out of CH$_4$ become less efficient at warmer temperatures. The carbon chain abundances also decrease with  $T_{\rm init}$ and $T_{\rm min}$, and the abundances of C$_4$H and C$_3$H$_2$ do not increase inwards when $T_{\rm init} \ge 20$ K and $T_{\rm min}\ge 20$ K. As an example,  $R-$abundance plot of CH$_4$ and carbon chains in the model with $T_{\rm min}=20$ K is depicted in Figure \ref{Tinit_WCCC} (c). 

\cite{sakai13} argued that the CH$_4$ abundance relative to H$_2$ should be higher than several $10^{-7}$ for the WCCC to be active, i.e. for CH$_4$ to be the major reactant for C$^+$ competing with OH and other molecules. In the sublimation region of CH$_4$ in our fiducial model, the reaction of C$^+$ + CH$_4$ $\rightarrow$ C$_2$H$_3^+$ + H competes with C$^+$ + H$_2$ $\rightarrow$ CH$_2^+$ and C$^+$ + NH$_3 \rightarrow$ H$_2$CN$^+$ + H, although the rates are higher for the latter two. The reaction of C + CH$_3 \rightarrow$ C$_2$H$_2$ + H also contributes the WCCC, where a fraction of CH$_4$ contributes to form C atom and CH$_3$, as well. 
While the model details are different between the present work and \cite{sakai13}, we can see in Figure \ref{Tinit_WCCC} that the gaseous CH$_4$ abundance of $10^{-7}$ would roughly be a condition for WCCC. The abundances of C$_2$H, C$_4$H, and C$_3$H$_2$ correlate with the sublimated abundance of CH$_4$.

In Figure \ref{Tinit_WCCC} (a)(b), the dash-dotted orange lines depict the CH$_4$ ice abundance at the end of the static phase, while the (non-vertical) solid orange lines depict the gaseous CH$_4$ abundance at the final timestep ($R=30.6$ au). Except for the models with $T_{\rm init} > 20$ K, the solid line overlaps with the dash-dotted line, which means that the total abundance of CH$_4$ is mostly determined by the CH$_4$ ice abundance in the static phase. In the models with $T_{\rm init} = 25$ K, the CH$_4$ abundance increases on the grain surfaces in the early collapse phase. We also note that the (non-vertical) solid lines are well above the upper values of the orange vertical solid lines, which means that  only a fraction of icy CH$_4$ sublimates in the WCCC region, except in the model of $T_{\rm init}\ge 20$ K.
The trapping and layering of molecules in ice determine what fraction of CH$_4$ can sublimates at $T\sim 25$ K and contributes to the WCCC.


\subsection{Dependence on UV extinction by ambient clouds}

\begin{figure}[ht!]
\plotone{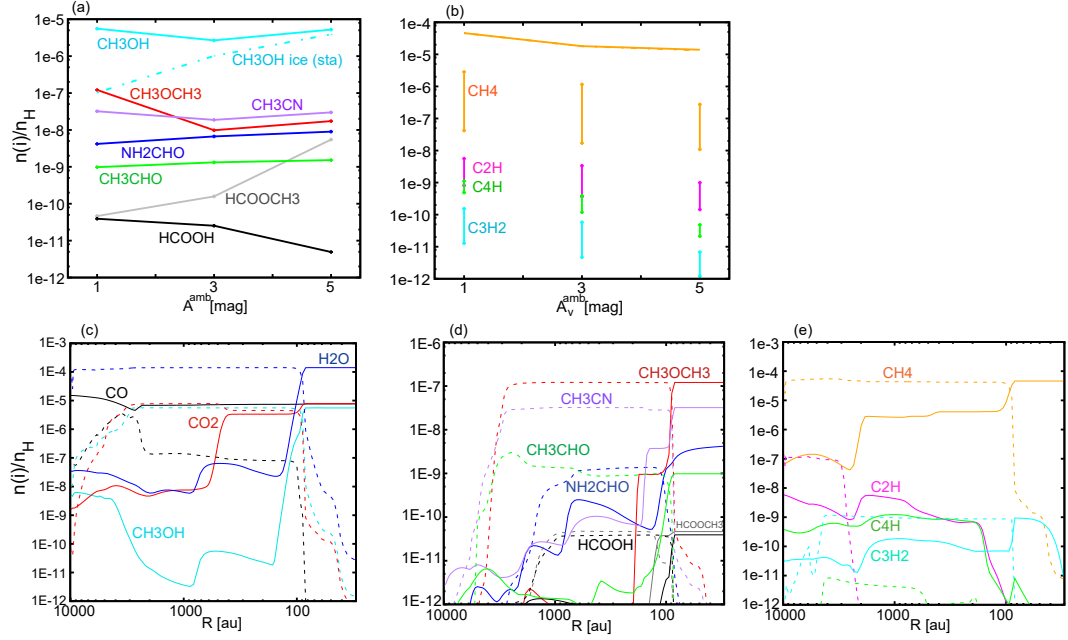}
\caption{(a)  Gas phase abundances of COMs at the final time step in models with the visual extinction by ambient clouds $A_{\rm v}^{\rm amb}=1,3,$ and 5 mag. The dash-dotted cyan lines depict the CH$_3$OH ice abundance at the end of the static phase.
(b) The maximum and minimum abundances of gaseous CH$_4$ (orange), C$_2$H (magenta), C$_4$H (green), and C$_3$H$_2$ (cyan) around the CH$_4$ sublimation region ($R=1455-2850$ au) as a function of $A_{\rm v}^{\rm amb}$. The upper solid orange line depicts the CH$_4$ gas abundance at the final timestep (i.e. $R=30.6$ au), while the dash-dotted orange line (which overlaps with the solid line) depicts the CH$_4$ ice abundant at the end of the static phase.
(c-e) $R-$abundance plot in the collapse phase in the model of with $A_{\rm v}^{\rm amb}=1$ mag. 
\label{Av_COM_WCCC}}
\end{figure}

In the fiducial model, we assumed that the initial prestellar core is embedded in ambient gas, which attenuate the interstellar UV radiation field by $A_{\rm v}^{\rm amb}=3$ mag (Figure \ref{fig_schem}). We calculated two additional models with $A_{\rm v}^{\rm amb}=1$ mag and 5 mag; i.e. the visual extinction in the static phase is 2.51 mag and 6.51 mag in the fluid parcel.
Figure \ref{Av_COM_WCCC} shows the gaseous COM abundances at the final time step (panel a) and the minimum and maximum abundances of CH$_4$, C$_2$H, C$_3$H$_2$, and C$_4$H around the CH$_4$ sublimation region (panel b) as a function of $A_{\rm v}^{\rm amb}$. A naive expectation is that in the prestellar core with lower $A_{\rm v}^{\rm amb}$, CH$_3$OH would be less abundant, while CH$_4$ would be more abundant, which results in less abundant COMs and more active WCCC in the protostellar phase. While CH$_4$ abundance is indeed high and WCCC is more active in the models with lower  $A_{\rm v}^{\rm amb}$ (Figure \ref{Av_COM_WCCC} b), the dependence of COM abundances on  $A_{\rm v}^{\rm amb}$ is not so simple.

Figure \ref{Av_COM_WCCC} (c)-(e) shows the $R-$abundance plot of the infalling fluid parcel in the model with  $A_{\rm v}^{\rm amb}=1$ mag. In the static phase, photolysis makes CH$_3$OH and other COMs less abundant both in the gas phase and ice mantle in the model with  $A_{\rm v}^{\rm amb}=1$ mag than in the fiducial model. Their abundances, however, become as high as in the fiducial model, once the visual extinction gets high enough in the collapse phase. Even though CO ice abundance is lower than in the fiducial model, CH$_3$OH can be formed via CH$_3$ + OH. 

In the model with $A_{\rm v}^{\rm amb}=5$ mag, HCOOH is less abundant, while HCOOCH$_3$ is more abundant than in the fiducial model. 
In the fiducial model, HCOOH is formed around $R\sim 3\times 10^3$ au in the ice mantle via CO + OH $\rightarrow$ COOH and COOH + H $\rightarrow$ HCOOH. The abundance of OH radical in the ice mantle is lower in the model with $A_{\rm v}^{\rm amb}=5$ mag.
HCOOCH$_3$ is formed via HCO + CH$_3$O
in the ice mantle at $R\sim 1\times 10^3$ au in the model with $A_{\rm v}^{\rm amb}=5$ mag. Both of the reactants are formed in the reaction network of CH$_3$OH (i.e. photo-dissociation of CH$_3$OH and CH$_3$O + H$_2$CO $\rightarrow$ CH$_3$OH + HCO), which is more abundant in the ice mantle of  $A_{\rm v}^{\rm amb}=5$ mag than in the fiducial model.

\subsection{Dependence on the duration of static phase}

In the classical pseudo-time dependent models of molecular clouds, carbon is initially in the form of C$^+$, which is converted to C atom and then to CO. Carbon chains and CH$_4$ are formed from C atoms and thus reach the maximum abundance in the gas phase before carbon is fully converted to CO \citep[e.g.][]{suzuki92}. Then one would naively expect that if the star formation sets in earlier, the ice abundance ratios of CO/CH$_4$ and thus CH$_3$OH/CH$_4$ would be lower, which is favorable for WCCC rather than hot corinos.

\begin{figure}[ht!]
\plotone{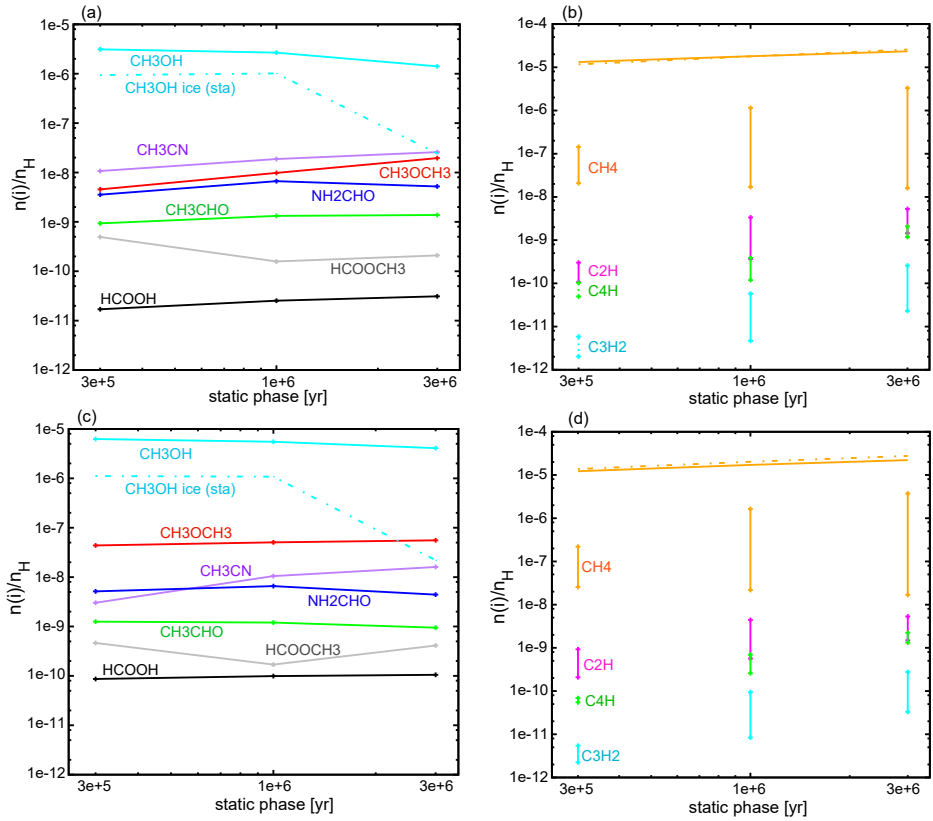}
\caption{(a)  Gas phase abundances of COMs at the final time step in models with the static phase of $t_{\rm sta}=
3\times 10^5$ yr, $1\times 10^6$ yr (fiducial), and $3\times 10^6$ yr. The dash-dotted cyan lines depict the CH$_3$OH ice abundance at the end of the static phase.
(b) The maximum and minimum abundances of gaseous CH$_4$ (orange), C$_2$H (magenta), C$_4$H (green), and C$_3$H$_2$ (cyan) around the CH$_4$ sublimation region ($R=1455-2850$ au) as a function of $t_{\rm sta}$. The upper solid orange line depicts the CH$_4$ gas abundance at the final timestep (i.e. $R=30.6$ au), while the dash-dotted orange line depicts the CH$_4$ ice abundant at the end of the static phase.
(c)-(d) Same as panel (a) and (b) but with the initial molecular abundance set by molecular formation in the converging flow.
\label{precol_COM_WCCC}}
\end{figure}

In order to test this expectation, we run models with various $t_{\rm sta}$;
Figure \ref{precol_COM_WCCC} shows the gaseous COM abundances at the final time step (panel a) and the minimum and maximum abundances of CH$_4$, C$_2$H, C$_3$H$_2$ and C$_4$H around the CH$_4$ sublimation region (panel b).
While the dependence of COM abundances on $t_{\rm sta}$ is rather weak, the abundance jump of carbon chains around CH$_4$ sublimation region is larger (i.e. WCCC is more active) in models with longer $t_{\rm sta}$. This counter-intuitive dependence is caused by the stability of CH$_4$ in the ice. Referring to Figure \ref{dist_fid}, we note that after the conversion of C atom to CO and CO ice, CH$_4$ ice abundance does not decrease, while gaseous CH$_4$ (and other carbon chains in the gas phase) does decrease. Actually, at $t\gtrsim 10^6$ yr in the static phase, CO is gradually converted to CH$_4$ via CO + He$^+$ $\rightarrow $ C$^+$ + O + He and subsequent reactions with H$_2$. The conversion of CO to CH$_4$ is also found in previous work, e.g. \cite{hassel08}, and is also responsible for the slight decrease of CH$_3$OH with $t_{\rm sta}$. 
We also note that in the models with $t_{\rm sta}=3\times 10^5$ yr, the abundances of C$_3$H$_2$ and C$_4$H decrease inwards around the CH$_4$ sublimation radius. Their abundances at the outermost radius ($\sim 10^4$ au) are higher than those in our fiducial model by about one order of magnitude. In other words, the remnant carbon chains from the prestellar phase dominates over production via WCCC.

So far, we assumed that species are in the form of atoms or atomic ions except for hydrogen, which is in H$_2$, at the start of static phase with $n_{\rm H}\sim 2\times 10^4$ cm$^{-3}$ and $A_{\rm v}\gtrsim 2$ mag. Such an initial abundance is often assumed, e.g. in pseudo-time dependent models and collapsing core models, because H$_2$ is self-shielded to be firstly formed, which is essential for the gas-phase two-body reactions to proceed. But it is obviously a simplified initial condition. Observations indicate that molecular formation both in the gas and ice phases starts in lower densities \citep[e.g.][]{whittet09, snow06}. Theoretical work shows that molecular clouds are formed via converging flows of HI gas \citep[e.g.][]{inoue12}, and the ice formation in the post-shock gas of converging flow reproduces important features of interstellar ices: D/H ratio of water ice and inhomogeneous distribution of polar (i.e. water) and apolar species within the ice mantle \citep{furuya15}. Since we aim to investigate the effect of ice composition on protostellar chemistry, we calculated another set of model, in which the initial molecular abundance of the static phase is set by solving the gas-grain chemistry in the converging flow. We solve the molecular evolution in the post-shock gas in the same 1D steady-state shock model as \cite{furuya15} \citep{bergin04, hassel10}, and adopt the molecular abundance when $A_{\rm v}$ reaches 1.2 mag. This choice of visual extinction is rather arbitrary,  but it roughly corresponds to the time when both gas-phase CO and water ice abundances reach the abundance of $10^{-4}$; CO abundance starts to decline due to the conversion to CH$_4$ at larger $A_{\rm v}$. The gas and ice are then put in the prestellar core, which is kept static for $t_{\rm sta}=3\times 10^5$ yr, $1\times 10^6$ yr or $3\times 10^6$ yr.

Figure \ref{precol_COM_WCCC} ($c, d$) shows the gaseous COM abundances and the minimum and maximum abundances of CH$_4$ and carbon chains as in Figure \ref{precol_COM_WCCC} ($a, b$), but with the initial abundance set in the converging flow. We can see that the COM abundances, activity of WCCC, and their dependence on  $t_{\rm sta}$ are similar to those in panels ($a$, $b$). We note that the abundance of CH$_3$OCH$_3$ is higher in the model with converging flow. It is most abundantly formed in the inner-most layer of the ice mantle, the molecular abundance of which reflects the early evolution. A major characteristics of ice mantle composition set by the converging flow is that the inner-most layer is CO-poor, which results in relatively high abundance of OH. In our fiducial model, CO ice is more uniformly distributed in bulk ice layers, and reacts with OH to form CO$_2$, when the thermal diffusion becomes efficient. CH$_3$OCH$_3$ is formed via CH$_3$ + CH$_3$O. In the OH-rich ice layer, CH$_3$ reacts with OH to reform CH$_3$OH, which is photodissociated to CH$_3$O. In the fiducial model, on the other hand, CH$_3$ reacts mainly with NH$_2$ to form CH$_3$NH$_2$. Thus CH$_3$CN is more abundant in our fiducial model, while CH$_3$OCH$_3$ is more abundant with the ice mantle set by the converging flow.

\subsection{a grid of models}

So far we varied one of the parameters of temperature, the visual extinction of ambient gas ($A_{\rm v}^{\rm amb}$), or the duration of the static phase ($t_{\rm sta}$). COMs abundances and WCCC activity in the protostellar phase are found to be most sensitive to the minimum temperature among these parameters.  In this subsection, we investigate how this sensitivity depends on $A_{\rm v}^{\rm amb}$ and $t_{\rm sta}$. 

\begin{figure}[ht!]
\plotone{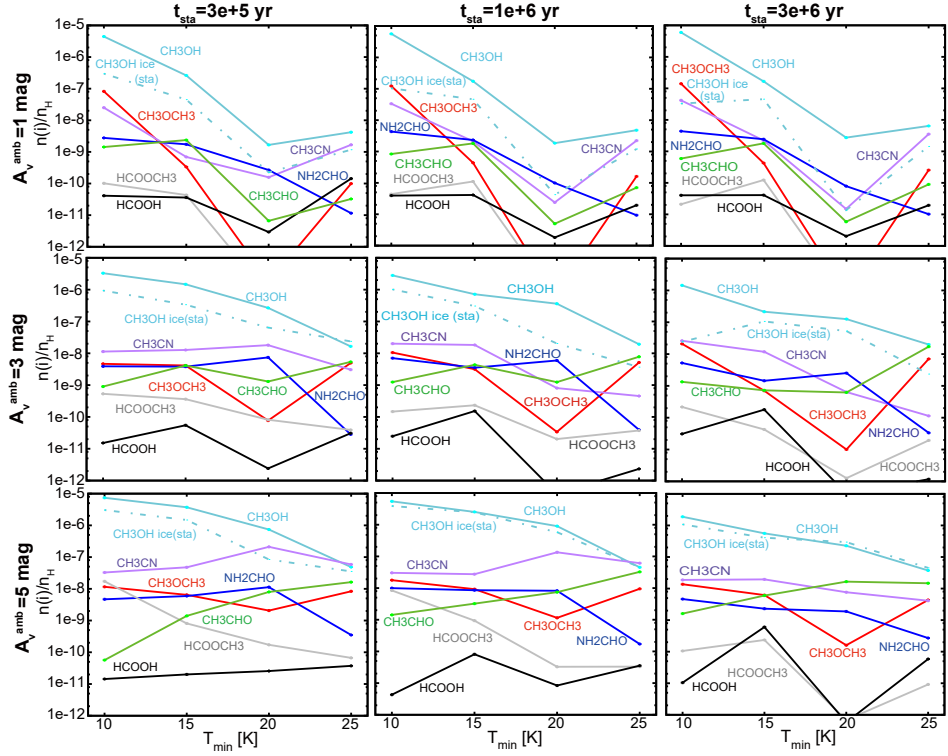}
\caption{Gas phase abundances of COMs at the final time step (solid lines) and CH$_3$OH ice abundance at the end of the static phase (dash-dotted lines) as a function of $T_{\rm min}$ in the grid of models. 
\label{two_param_COM}}
\end{figure}

Figure \ref{two_param_COM} shows the COM abundances as a function of $T_{\rm min}$ in models with $A_{\rm v}=1$ mag, 3 mag, and 5 mag and $t_{\rm sta}=3\times 10^5$ yr, $1\times 10^6$ yr, and $3\times 10^6$ yr. We can see that the dependence of COM abundances on $T_{\rm min}$ is more significant in models with lower $A_{\rm v}^{\rm amb}$. In these models, photodissociation of molecules, including COMs, is more effective. While the photodissociation produces radicals, which can recombine to reform COMs, the rates of their diffusion in ice mantle and sublimation (i.e. loss to the gas phase) are very sensitive to temperatures. Photodissociation thus enhances the dependence of COM abundances on temperatures. The dependence of some COM abundances on $T_{\rm min}$ is also stronger in models with longer static phase.
In the middle row (Av=3mag) in Figure \ref{two_param_COM}, for example, the CH$_3$CN abundance varies more than two orders of magnitudes in the models with $t_{\rm sta}=3 \times 10^6$ yr, while the variation is within an order of magnitude in the models with $t_{\rm sta}=3 \times 10^5$ yr.
When a gas-grain reaction network is kept at a constant temperature, specific molecules and radicals accumulate in ices: e.g. species that can freeze-out and/or species produced by the recombination of radical that can thermally diffuse. If the duration of the constant temperature is longer, the accumulation of those specific species becomes more significant, which results in the higher dependence of protostellar core chemistry on $T_{\rm min}$.
Comparing the solid and dash-dotted cyan lines, we note that a significant amount of CH$_3$OH is formed after the onset of collapse in models with low $A_{\rm v}^{\rm amb}$ ($\lesssim 3$ mag). The abundance ratio of COMs to gaseous CH$_3$OH, as well as that to CH$_3$OH ice at the end of static phase, vary significantly with $T_{\rm min}$.

\begin{figure}[ht!]
\plotone{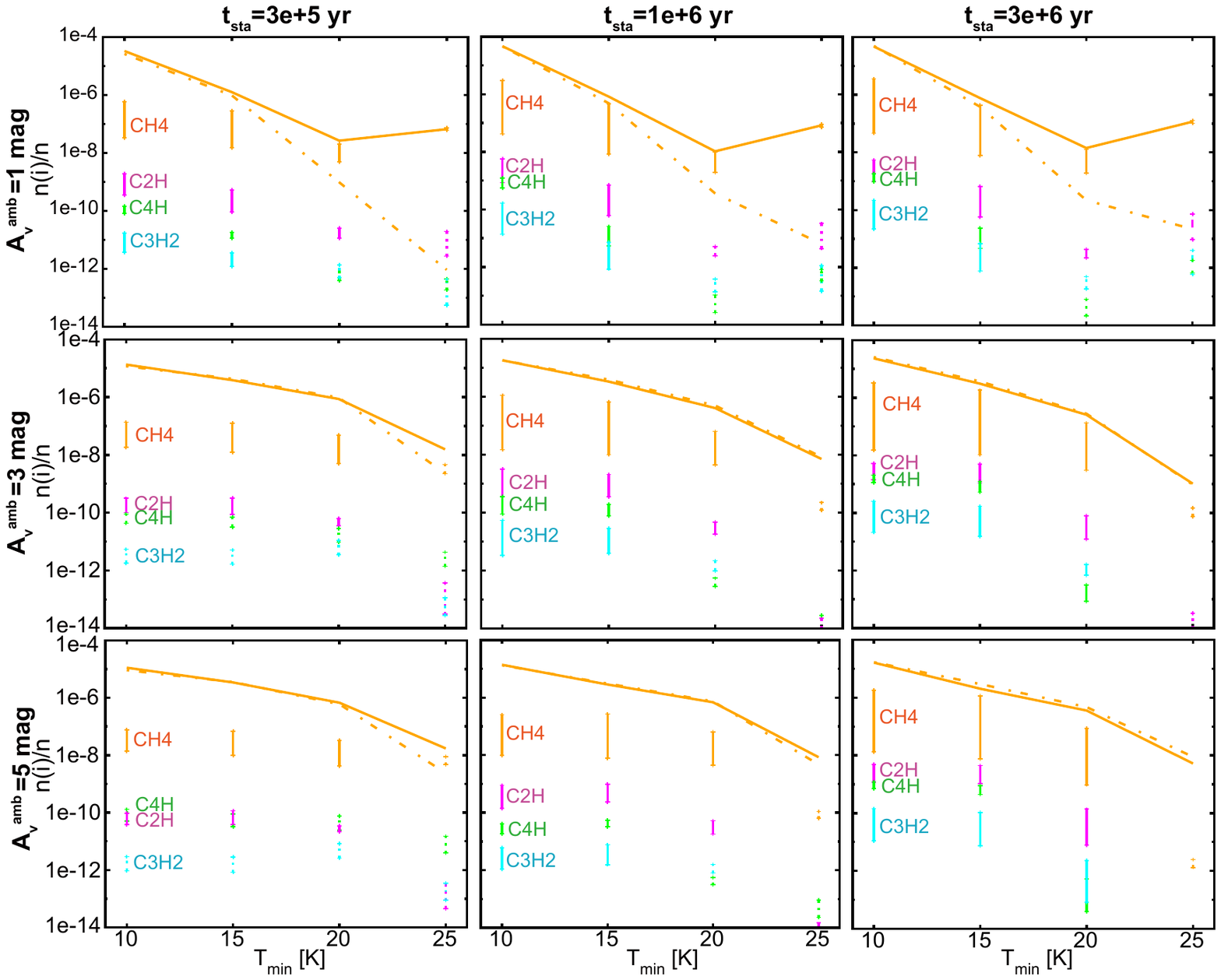}
\caption{Gas phase abundances of CH$_4$ and carbon chains around the CH$_4$ sublimation region as a function of $T_{\rm min}$ in the grid of models. The upper solid orange lines depict the CH$_4$ gas abundance at the final timestep (i.e. $R=30.6$ au), while the dash-dotted oranges line depict the CH$_4$ ice abundant at the end of the static phase.
\label{two_param_WCCC}}
\end{figure}

The WCCC activity in the same set of models are shown in Figure \ref{two_param_WCCC}.
The WCCC is activated when the sublimated CH$_4$ abundance is higher than $\sim 10^{-7}$. In models with $A_{\rm v}^{\rm amb}= 3$ mag and 5 mag, the CH$_4$ ice abundance at the end of static phase (and the gaseous CH$_4$ at the final timestep) does not sensitively depend on $t_{\rm sta}$. The sublimated CH$_4$ abundance around WCCC region (i.e. $T\sim 25$ K) is, however, higher in models with larger $t_{\rm sta}$. Since the conversion of CO to CH$_4$ becomes efficient later in the static phase, the CH$_4$ ice in the surface layer, which is subject to the immediate sublimation at $\sim 25$ K, is more abundant in models with larger $t_{\rm sta}$. WCCC is thus more active in models with longer $t_{\rm sta}$. In the models with $A_{\rm v}^{\rm amb}=1$ mag and $T_{\rm min}\gtrsim 20$ K, CH$_4$ ice abundance is very low at the end of static phase, and thus WCCC is not active. In these models, a large fraction of CH$_4$ is formed after the onset of collapse by the gas-phase reactions starting from C$^+$ + H$_2$ $\rightarrow$ CH$_2^+$. WCCC is activated in models with $T_{\rm min}\lesssim 15$ K, even if the ambient visual extinction is low (i.e. $A_{\rm v}^{\rm amb}=1$ mag).

\section{Discussion}
\subsection{comparison with previous work}

The abundances of carbon chains and COMs in our fiducial model are different from those in our previous work \citep{aikawa08, aikawa12}, since we adopt the multi-layered ice mantle model, rather than the two-phase model as described in \S 3.1. Entrapment of CH$_4$ suppresses the WCCC, while the entrapment of radicals enhances the formation of COMs \citep{lu18}.


\cite{acharyya18} investigated the hot corino chemistry in Large Magellanic Cloud and Small Magellanic Cloud by calculating the two-phase model of gas-grain chemistry in the cold collapse stage and warm-up stage. The dependence of peak molecular abundances on the temperature in the collapse stage is investigated, as well. The peak abundances of COMs are basically lower in models with higher temperature in the collapse phase, which is qualitatively consistent with our results on $T_{\rm min}$ (\S 3.2). The COM abundances in our model with $T_{\rm min}=25$ K, however, tends to be much higher than those in \cite{acharyya18}. It would be mostly due to the effect of multi-layered ice mantle model, which can keep various radicals even at warm temperatures. Our grid of models also suggest that the COM abundances depend more sensitively on $T_{\rm min}$ when $A_{\rm v}^{\rm amb}$ is lower. It is in line with the recent observations, which show large abundance variations of COMs among cores in LMC and SMC
\citep{shimonishi16, sewilo18, shimonishi18, shimonishi20}, since the visual extinction is lower in those low-metalicity galaxies compared with that in our Galaxy.

\cite{vidal19} calculated the three-phase model in 110 models of star-forming core of \cite{vaytet17}. They statistically analyzed the correlation of molecular abundances and physical model parameters, and found that CH$_3$CN abundance correlates positively with the initial core temperature. Since they plot the final molecular abundance in all the fluid parcels of 110 models as a function of the initial temperature, CH$_3$CN abundance is scattered over one order of magnitude or more at each initial temperature bin. The dependence of the CH$_3$CN abundance (i.e. the maximum or mean value) on the initial temperature is similar to the dependence of CH$_3$CN on $T_{\rm init}$ in our model (Figure \ref{Tinit}); the dependence is weak at $T_{\rm init} =10-20$ K, while the abundance is about one-order of magnitude higher at $T_{\rm init}=25$ K. They concluded that the positive correlation is caused by enhanced diffusion of CN and CH$_3$ in the ice mantle. We speculate that this explanation is too simplified; other COM abundances would show a positive correlation with the initial core temperature, if the diffusion rate is the key. As we discussed in \S 3, the formation paths of COMs vary with the initial temperature; new formation paths open in the model with $T_{\rm init}=25$ K.

We adopted the chemical reaction network of \cite{garrod13}, which investigated formation of COMs, especially  glycine, in warm-up models mimicking the star-forming core. Their model consists of the cold collapse phase and warm-up phase. In the former, the initial gas density increases from $n_{\rm H}=3 \times 10^3$ cm$^{-3}$ to $1\times 10^7$ cm$^{-3}$ in $\sim 10^6$ yr. The dust temperature and visual extinction $A_{\rm v}$ are initially 16 K and 2 mag, respectively. The temperature decreases as $A_{\rm v}$ increases, reaching the minimum value of 8 K. In the warm-up phase, temperature increases from 8 K to 400 K, while the density is kept constant ($1\times 10^7$ cm$^{-3}$). Three models are calculated with the warm-up timescale of $7.12 \times 10^4$ yr (fast), $2.85\times 10^5$ yr (medium), and $1.43 \times 10^6$ yr (slow). Comparing the timescale of temperature rise from 20 K to 100 K, our model is similar to the fast or medium model of \cite{garrod13}, in which the peak abundances of gaseous CH$_3$OCH$_3$, CH$_3$CN, and CH$_3$CHO are $(3-5) \times 10^{-8}$, $(2-5) \times 10^{-9}$, and $(3-9)\times 10^{-9}$, respectively. Despite the differences in the physical model and chemical model (the three-phase model with swapping versus the multi-layered ice mantle model without swapping), these COM abundances are similar to our results. Our abundances of NH$_2$CHO and HCOOH at the final timestep are smaller than those in \cite{garrod13} by more than an order of magnitude, since we deleted some reactions relevant for their formation (\S 2.2) (see also \S 4.2).



\subsection{uncertainties in the reaction network of COMs}

Besides the treatment of grain surface chemistry (e.g. two-phase, three-phase, and multi-phase model) and physical model of core formation and evolution, there are uncertainties in chemical reaction network. One of the major uncertainties in the hot corino chemistry is the branching ratio of radical reactions in the ice, as it is difficult to directly measure in laboratory experiments. Quantum chemical calculations have been useful to estimate such branching ratios, but are often not straightforward, since the interaction with the grain surface and neighboring icy species need to be included \citep[e.g.][]{kayanuma19}.
While the main aim of the present work is to investigate how the WCCC and hot corino chemistry as a whole depend on the physical conditions in prestellar pase, rather than the dependence of each COM species, it is useful to check the effect of uncertainties in reaction network on our model results.

Specifically we modified the branching ratio of two grain surface reactions: NH$_2$ + H$_2$CO and HCO + CH$_3$.  In the gas phase, \cite{barone15} found that the activation barrier of NH$_2$ + H$_2$CO $\rightarrow$ NH$_2$CHO + H is very low (26.9 K), which we adopt in our fiducial model. For the reaction of NH$_2$ + H$_2$CO in ice, on the other hand, we assumed the products to be NH$_3$ + HCO with the activation barrier of 2360 K, as assumed in the original network of \cite{garrod13} to be conservative (see also discussions in \cite{fedoseev16}).
In order to check the effect of uncertainty of this reaction, here we assume that NH$_2$CHO and H are formed without activation barrier in ice. As for the icy reaction of HCO + CH$_3$, we assume the products to be CH$_4$ + CO in our fiducial model referring to  \cite{enrique-romero16}.  Recently \cite{enrique-romero20} re-investigated this reaction with the broken-symmetry approach to find that the formation of CH$_3$CHO proceeds without barrier, while the direct H transfer to form CO + CH$_4$ can be a competitive channel. We thus assume 1:1 branching ratio for the product channels of CH$_3$CHO and CO + CH$_4$.

\begin{figure}[ht!]
\includegraphics[scale=0.7]{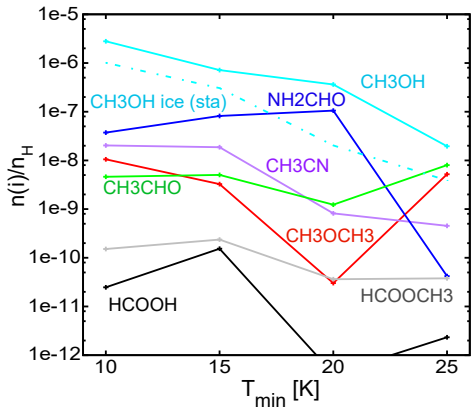}
\caption{Gas phase abundances of COMs at the final time step ($R=30.6$ au) in models with the minimum temperature of 10 K, 15 K, 20 K, and 25 K with the modified branching ratios of NH$_2$ + H$_2$CO and HCO + CH$_3$. The dash-dotted cyan lines depict the CH$_3$OH ice abundance at the end of the static phase.
\label{mod_COM}}
\end{figure}

Figure \ref{mod_COM} shows the COM abundances at the final timestep ($R=30.6$ au) in the models with modified branching ratios with $T_{\rm min}$ of 10 K, 15 K, 20 K, and 25 K. Compared with Figure \ref{Tinit} (b), the abundance of CH$_3$CHO is enhanced by a factor of 3.7 in the model with $T_{\rm min}=10$ K, while that of NH$_2$CHO is enhanced
by a factor of $5.3-23$ in the models with $T_{\rm min} \le 20$ K. It is as expected, since we modified the branching ratios to be favorable for the formation of these molecules. It also suggests that other reaction paths dominate in their formation at higher temperatures. In order to discuss the dependence of each COM abundance on prestellar temperatures, it is essential to refine the chemical reaction network by the laboratory experiments and quantum chemical calculations.

\subsection{comparison with observations}

We found that WCCC, i.e. the formation of carbon chains around the CH$_4$ sublimation region, is more active and their abundances are higher in models with lower $T_{\rm init}$ and $T_{\rm min}$, lower $A_{\rm v}^{\rm amb}$, and longer $t_{\rm sta}$. The abundance of C$_4$H in L1527 and IRAS 15398, which are prototypical WCCC sources, is $\gtrsim 10^{-9}$, while it is $\sim 10^{-11}$ in a prototypical hot corino IRAS 16293 \citep{sakai09a}. This range of C$_4$H abundance is covered by the models in the present work.
The molecular D/H ratio could be a key to discriminate which parameter, $T_{\rm init}$ ($T_{\rm min}$), $A_{\rm v}^{\rm amb}$, or $t_{\rm sta}$, is responsible for the variation of carbon chain abundances in protostellar cores; low $T_{\rm init}$ and long $t_{\rm sta}$ would enhance D/H ratio, while low D/H ratio is expected for low $A_{\rm v}^{\rm amb}$.
In L1527, the column density ratio of c-C$_3$HD/c-C$_3$H$_2$ is 4.4 \% \citep{yoshida19} (see also \cite{sakai09b}), while the ratio is observed to be 14 \% towards IRAS 16293 \citep{majumdar17}, which may suggest that the variation is caused by the visual extinction. 

The dependence of COM abundances on the static-phase conditions are more complex, since there are various formation paths of COMs, and since their efficiency depends on the composition of each layer of ice mantle. Among the parameters investigated, $T_{\rm min}$ is the most effective. 
While CH$_3$OH basically decreases with $T_{\rm min}$, CH$_3$OCH$_3$, for example, is least abundant in the model with $T_{\rm min}=20$ K, and is more abundant in the model with $T_{\rm min}=25$ K, in which COMs can be formed from large hydrocarbons.
It should be noted that the observations often evaluate the relative COM abundances to CH$_3$OH. In our models, the COM abundances relative to CH$_3$OH tend to increase with $T_{\rm min}$ and $T_{\rm init}$ as CH$_3$OH decreases, although either CH$_3$OH or other COMs would not be detected if their abundances are too low.
Recently, \cite{oya19} observed Class I protostellar source Elias 29 to find both COMs and carbon chains are deficient; the abundances of C$_2$H and c-C$_3$H$_2$ are $\lesssim 10^{-11}$ and those of HCOOCH$_3$ and CH$_3$OCH$_3$ are $\lesssim 10^{-9}$. Considering the uncertainties of the upper limits, these abundances are consistent in our model with $T_{\rm min}= 20$ K.



In our models with  higher $T_{\rm init}$, higher $A_{\rm v}^{\rm amb}$, or shorter $t_{\rm sta}$, WCCC is less active, while COMs can be abundantly formed; these models can explain the deficiency of carbon chains in hot corinos. 
The deficiency of COMs in WCCC sources is, on the other hand, hard to reproduce by simply varying the static-phase conditions; COM abundances do not monotonically decrease with a specific parameter, and COMs are abundantly formed in models with active WCCC. While COM abundances are low in the model with $T_{\rm min}=20$ K, WCCC is not active, either. 
Alternatively, the deficiency of COM emissions could be due to temperature distributions in the WCCC sources. The sublimation temperature of COMs is typically 100 K. The size of the hot corino, i.e. the radius of the COM sublimation region, is typically $\lesssim 100$ au. If the central protostar is less luminous, the sublimation region could be significantly smaller; the COM emission lines could be weakened by the beam dilution and high dust opacity, since the (column) density is higher at smaller radii. 
Indeed, the luminosity of the prototypical hot corino sources (e.g. 9.1 $L_{\odot}$ in NGC 1333 IRAS4A and 22 $L_{\odot}$ in IRAS 16293-2422 ) tend to be higher than that of WCCC sources; e.g. 1.9 $L_{\odot}$ and 1.8 $L_{\odot}$ for L1527 and IRAS 15398, respectively \citep{froebrich05, crimier10, kristensen12, karska13, jorgensen13}. One notable exception is B335; while its luminosity is as low as 0.72 $L_{\odot}$, COMs emission is detected within a few 10 au at the core center, where the fractional abundances of COMs are  comparable to those in the prototypical hot corino IRAS 16293 \citep{imai16}. A possible explanation would be that B335 experienced temporal outburst, which sublimated COMs, but is currently back to its quiescent phase. COMs are still in the gas phase, if the time after the outburst
is less than the re-freeze-out timescale, i.e. $\sim 10^3 (10^7$ cm$^{-3}$/$n_{\rm H}$) yrs. 

\begin{figure}[ht!]
\includegraphics[scale=0.4, angle=270]{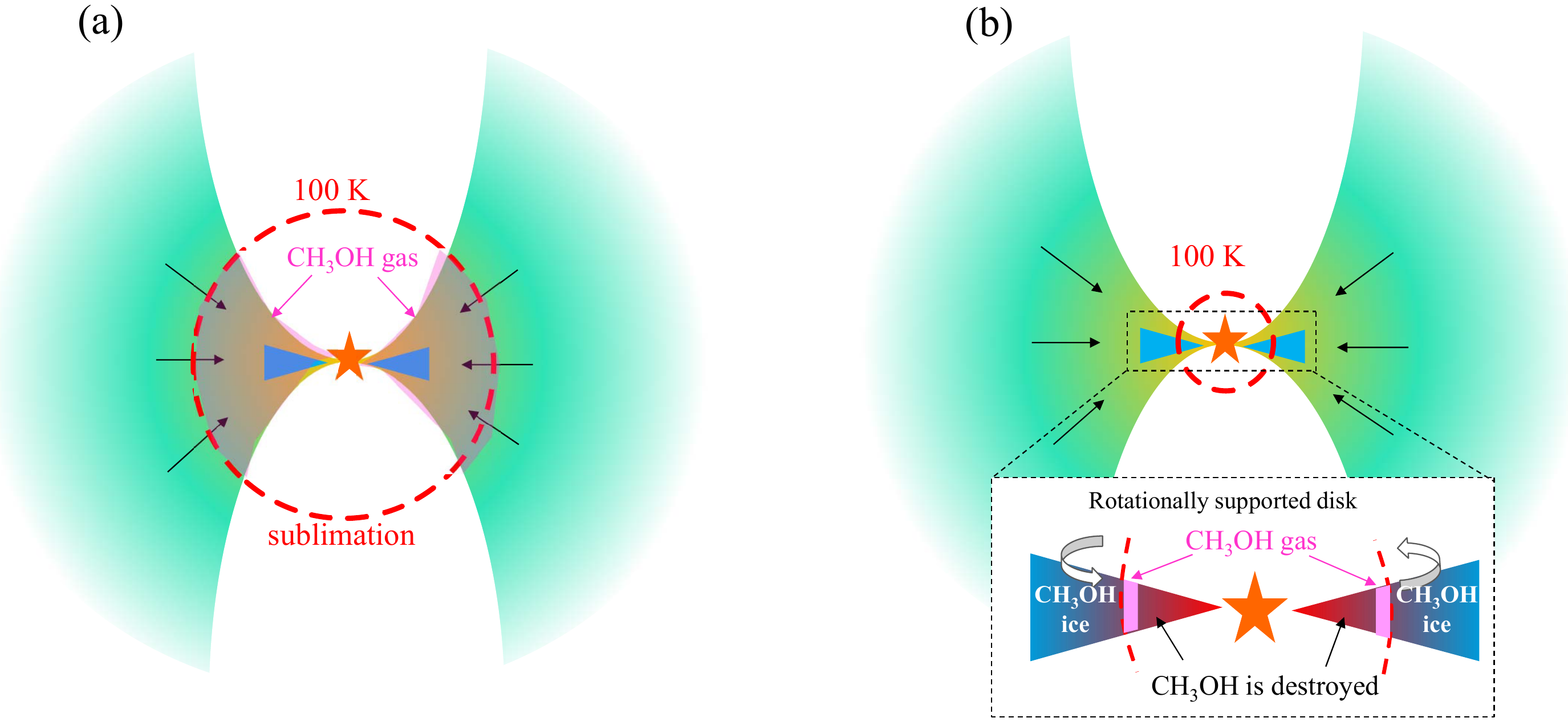}
\caption{Schematic view of CH$_3$OH gas distribution in (a) a hot corino source with the $T\sim 100$ K region located in the infalling envelope and (b) a protostellar core with the $T\sim 100$ K region located in the rotationally-supported disk.
\label{discussion}}
\end{figure}

We note that the kinetic structure inside a few hundreds au is as relevant as temperature distributions.  
Once CH$_3$OH and other COMs are sublimated to the gas-phase, they are destroyed by gas-phase reactions within several $10^4$ yrs \citep{charnley92, nomura09, taquet16}. The spatial extent of gaseous COMs is thus estimated to be the product of destruction timescale and radial velocity of the gas. When the $T\sim 100$ K region is located in the envelope, we can naively expect constant gaseous abundance of COMs inside this region, since infall is faster than COM destruction in the gas phase (Figure \ref{discussion} a). If the $T\sim 100$ K region is inside the rotationally-supported disk, on the other hand, the spatial extent of gas-phase COMs would be very narrow, limited by the competition between gas-phase destruction of COMs and slow radial migration of gas in the disk (Figure \ref{discussion} b). While the spatially resolved observation of hot corino is challenging, \cite{imai19} recently resolved the COM emission in B335, and showed that the variation between velocity gradients of COM emissions are well explained by the model of infalling and rotating gas, rather than Keplerian motion. Observations with high spatial resolution (e.g. $\sim 0.1$ \arcsec)
or in lower frequency band, in which the dust opacity is lower, are desirable to investigate the physical structure such as temperature and density distributions, and the radius of the forming disk.



\section{Summary}
We investigated the dependence of WCCC and hot corino chemistry on the physical parameters in the static phase before the onset of collapse: the initial and minimum temperatures ($T_{\rm init}$ and $T_{\rm min}$), visual extinction of ambient gas ($A_{\rm v}^{\rm amb}$), and the duration of the static phase ($t_{\rm sta}$). Our findings are as follows.
\begin{itemize}
\item {Among the parameters, $T_{\rm min}$ is the most effective on COM abundances. CH$_3$OH and some other COMs tend to decrease with increasing $T_{\rm min}$, since freeze-out of molecules and hydrogenation on grain surfaces are less efficient at warm temperatures. But molecules with higher sublimation temperatures can still be adsorbed onto grains, and there are various formation paths of COMs, some of which become efficient at warm temperature. The abundance of CH$_3$OCH$_3$, for example, is higher in the model with $T_{\rm min}=25$ K than that with $T_{\rm min}=20$ K. Dependence of COM abundances on $T_{\rm init}$ is weaker, since various molecules, including CO, can be frozen and hydrogenated on grain surfaces during the cold phase right after the onset of collapse. }
\item{The gaseous CH$_3$OH abundance in the central hot region ($\ge 100$ K) monotonically decreases with increasing $T_{\rm init}$ and $T_{\rm min}$, except for the models with low visual extinction ($A_{\rm v}^{\rm amb}=1$ mag). Dependence of other COM abundances on $T_{\rm init}$ and $T_{\rm min}$ is not monotonic, as described above. The relative abundance of COMs to CH$_3$OH, which is often derived and discussed in the observational studies, could then be higher in cloud cores with higher $T_{\rm init}$ and $T_{\rm min}$, }
\item{WCCC is less active and carbon-chain species are less abundant in models with higher $T_{\rm init}$ or $T_{\rm min}$, since both the grain-surface formation and freeze-out of CH$_4$ become less effective at warm temperatures. Warm temperature also enhances the conversion of CO to CO$_2$ on grain surfaces, which reduces gaseous CO and the production of C$^+$ via CO + He$^+$ $\rightarrow$ C$^+$ + O + He. }
\item{While CH$_4$ and carbon chains are more abundant in the models with lower $A_{\rm v}^{\rm amb}$, the dependence of COM abundances on  $A_{\rm v}^{\rm amb}$ is more complex. Even though photolysis makes CH$_3$OH and other COMs less abundant in the static phase of lower $A_{\rm v}^{\rm amb}$ model, they can be formed via various reactions in ice mantle, e.g. CH$_3$ + OH $\rightarrow$ CH$_3$OH, once the collapse starts and $A_{\rm v}$ increases.}
\item{When the duration of static phase $t_{\rm sta}$ is varied from $3\times 10^5$ yr to $3\times 10^6$ yr, the COM abundances in the protostellar phase vary less than an order of magnitude. WCCC, on the other hand, is more active in the model with longer $t_{\rm sta}$, since CH$_4$ ice is stable and accumulate during the static phase. In the model with short $t_{\rm sta}$, the remnant of carbon chains from the prestellar phase dominates over those formed via WCCC. }
\item{We also calculated additional models in which the initial molecular abundances are set by considering the cloud formation via converging flow. The COM abundances and WCCC activities in the protostellar phase are basically similar to those in the fiducial model.
A notable difference is that the ice mantle has a larger chemical gradient, e.g. the inner most ice layer is deficient in CO ice. It enhances the abundances of OH radical and CH$_3$OCH$_3$ in the ice mantle in the protostellar phase compared with the fiducial model.}
\item{We calculated a grid of models to investigate how the dependence of COMs and WCCC on $T_{\rm min}$ vary with $A_{\rm v}^{\rm amb}$ and $t_{\rm sta}$.  Variation of COM abundances with $T_{\rm min}$ is enhanced in models with low $A_{\rm v}^{\rm amb}$.
The models with low $A_{\rm v}^{\rm amb}$ could also be relevant to the recent observations of hot cores in LMC and SMC, where significant variations are found in COM emission.
The dependence of some COM abundances on $T_{\rm min}$ is also stronger in models with longer $t_{\rm sta}$.
In the models with longer $t_{\rm sta}$, a larger fraction of CH$_4$ ice is in surface layers of ice mantle, which sublimates and activates WCCC at $T\sim 25$ K. }
\item{Our models show that the variation of $T_{\rm init}$ (or $T_{\rm min}$), $A_{\rm v}^{\rm amb}$, and $t_{\rm sta}$ can explain the chemical diversity between prototypical hot corinos, in which carbon-chains are deficient, and hybrid sources, toward which both COM and carbon chains are abundant.
A relatively low D/H ratio of carbon chain observed in a prototypical WCCC source L1527 may indicate that $A_{\rm v}^{\rm amb}$ is the key parameter.
Deficiency of COMs in prototypical WCCC sources is, however, hard to reproduce within our models; i.e. models with active WCCC have relatively abundant COMs. 
A possible explanation for the deficiency would be the small size of COM sublimation region (i.e. $\gtrsim 100$ K) and/or the kinetic structure there.
Brightness of the COM lines would be suppressed, if the COM sublimation region is significantly smaller than the beam size or within the rotationally-supported disk, in which the radial extent of gaseous COMs are limited by the competition between destruction in the gas-phase and slow radial accretion of gas. In case the COM sublimation region is small, the emission lines could also be hidden by the high dust opacity.
Observations with high spatial resolution are desirable to investigate the physical structure in the central regions of protostellar cores such as temperature and density distributions, and the radius of the forming disk.}
\end{itemize}

\acknowledgments

We thank R. T. Garrod for helpful discussions and sharing the chemical reaction network model. We thank the anonymous referee for his/her constructive comments.
This work is supported by Grant-in-Aid for Scientific Research (S) 18H05222, Grant-in-Aid for Young Scientists (B) 17K14245, and NAOJ ALMA Scientific Research Grant Numbers 2019-13B.

\bibliography{reference}

\begin{thebibliography}{}
\expandafter\ifx\csname natexlab\endcsname\relax\def\natexlab#1{#1}\fi
\providecommand{\url}[1]{\href{#1}{#1}}
\providecommand{\dodoi}[1]{doi:~\href{http://doi.org/#1}{\nolinkurl{#1}}}
\providecommand{\doeprint}[1]{\href{http://ascl.net/#1}{\nolinkurl{http://ascl.net/#1}}}
\providecommand{\doarXiv}[1]{\href{https://arxiv.org/abs/#1}{\nolinkurl{https://arxiv.org/abs/#1}}}

\bibitem[{{Acharyya} \& {Herbst}(2018)}]{acharyya18}
{Acharyya}, K., \& {Herbst}, E. 2018, \apj, 859, 51,
  \dodoi{10.3847/1538-4357/aabaf2}

\bibitem[{{Aikawa} {et~al.}(2001){Aikawa}, {Ohashi}, {Inutsuka}, {Herbst}, \&
  {Takakuwa}}]{aikawa01}
{Aikawa}, Y., {Ohashi}, N., {Inutsuka}, S.-i., {Herbst}, E., \& {Takakuwa}, S.
  2001, \apj, 552, 639, \dodoi{10.1086/320551}

\bibitem[{{Aikawa} {et~al.}(1997){Aikawa}, {Umebayashi}, {Nakano}, \&
  {Miyama}}]{aikawa97}
{Aikawa}, Y., {Umebayashi}, T., {Nakano}, T., \& {Miyama}, S.~M. 1997, \apjl,
  486, L51, \dodoi{10.1086/310837}

\bibitem[{{Aikawa} {et~al.}(2008){Aikawa}, {Wakelam}, {Garrod}, \&
  {Herbst}}]{aikawa08}
{Aikawa}, Y., {Wakelam}, V., {Garrod}, R.~T., \& {Herbst}, E. 2008, \apj, 674,
  984, \dodoi{10.1086/524096}

\bibitem[{{Aikawa} {et~al.}(2012){Aikawa}, {Wakelam}, {Hersant}, {Garrod}, \&
  {Herbst}}]{aikawa12}
{Aikawa}, Y., {Wakelam}, V., {Hersant}, F., {Garrod}, R.~T., \& {Herbst}, E.
  2012, \apj, 760, 40, \dodoi{10.1088/0004-637X/760/1/40}

\bibitem[{{Andr{\'e}} {et~al.}(2014){Andr{\'e}}, {Di Francesco},
  {Ward-Thompson}, {Inutsuka}, {Pudritz}, \& {Pineda}}]{andre14}
{Andr{\'e}}, P., {Di Francesco}, J., {Ward-Thompson}, D., {et~al.} 2014, in
  Protostars and Planets VI, ed. H.~{Beuther}, R.~S. {Klessen}, C.~P.
  {Dullemond}, \& T.~{Henning}, 27

\bibitem[{{Barone} {et~al.}(2015){Barone}, {Latouche}, {Skouteris}, {Vazart},
  {Balucani}, {Ceccarelli}, \& {Lefloch}}]{barone15}
{Barone}, V., {Latouche}, C., {Skouteris}, D., {et~al.} 2015, \mnras, 453, L31,
  \dodoi{10.1093/mnrasl/slv094}

\bibitem[{{Bergin} {et~al.}(2004){Bergin}, {Hartmann}, {Raymond}, \&
  {Ballesteros-Paredes}}]{bergin04}
{Bergin}, E.~A., {Hartmann}, L.~W., {Raymond}, J.~C., \& {Ballesteros-Paredes},
  J. 2004, \apj, 612, 921, \dodoi{10.1086/422578}

\bibitem[{{Boogert} {et~al.}(2008){Boogert}, {Pontoppidan}, {Knez}, {Lahuis},
  {Kessler-Silacci}, {van Dishoeck}, {Blake}, {Augereau}, {Bisschop},
  {Bottinelli}, {Brooke}, {Brown}, {Crapsi}, {Evans}, {Fraser}, {Geers},
  {Huard}, {J{\o}rgensen}, {{\"O}berg}, {Allen}, {Harvey}, {Koerner}, {Mundy},
  {Padgett}, {Sargent}, \& {Stapelfeldt}}]{boogert08}
{Boogert}, A.~C.~A., {Pontoppidan}, K.~M., {Knez}, C., {et~al.} 2008, \apj,
  678, 985, \dodoi{10.1086/533425}

\bibitem[{{Bottinelli} {et~al.}(2004{\natexlab{a}}){Bottinelli}, {Ceccarelli},
  {Lefloch}, {Williams}, {Castets}, {Caux}, {Cazaux}, {Maret}, {Parise}, \&
  {Tielens}}]{bottinelli04a}
{Bottinelli}, S., {Ceccarelli}, C., {Lefloch}, B., {et~al.} 2004{\natexlab{a}},
  \apj, 615, 354, \dodoi{10.1086/423952}

\bibitem[{{Bottinelli} {et~al.}(2004{\natexlab{b}}){Bottinelli}, {Ceccarelli},
  {Neri}, {Williams}, {Caux}, {Cazaux}, {Lefloch}, {Maret}, \&
  {Tielens}}]{bottinelli04b}
{Bottinelli}, S., {Ceccarelli}, C., {Neri}, R., {et~al.} 2004{\natexlab{b}},
  \apjl, 617, L69, \dodoi{10.1086/426964}

\bibitem[{{Bouvier} {et~al.}(2020){Bouvier}, {L{\'o}pez-Sepulcre},
  {Ceccarelli}, {Kahane}, {Imai}, {Sakai}, {Yamamoto}, \&
  {Dagdigian}}]{bouvier20}
{Bouvier}, M., {L{\'o}pez-Sepulcre}, A., {Ceccarelli}, C., {et~al.} 2020, \aap,
  636, A19, \dodoi{10.1051/0004-6361/201937164}

\bibitem[{{Caux} {et~al.}(2011){Caux}, {Kahane}, {Castets}, {Coutens},
  {Ceccarelli}, {Bacmann}, {Bisschop}, {Bottinelli}, {Comito}, {Helmich},
  {Lefloch}, {Parise}, {Schilke}, {Tielens}, {van Dishoeck}, {Vastel},
  {Wakelam}, \& {Walters}}]{caux11}
{Caux}, E., {Kahane}, C., {Castets}, A., {et~al.} 2011, \aap, 532, A23,
  \dodoi{10.1051/0004-6361/201015399}

\bibitem[{{Cazaux} {et~al.}(2003){Cazaux}, {Tielens}, {Ceccarelli}, {Castets},
  {Wakelam}, {Caux}, {Parise}, \& {Teyssier}}]{cazaux03}
{Cazaux}, S., {Tielens}, A.~G.~G.~M., {Ceccarelli}, C., {et~al.} 2003, \apjl,
  593, L51, \dodoi{10.1086/378038}

\bibitem[{{Ceccarelli} {et~al.}(2007){Ceccarelli}, {Caselli}, {Herbst},
  {Tielens}, \& {Caux}}]{ceccarelli07}
{Ceccarelli}, C., {Caselli}, P., {Herbst}, E., {Tielens}, A.~G.~G.~M., \&
  {Caux}, E. 2007, Protostars and Planets V, 47

\bibitem[{{Charnley} {et~al.}(1992){Charnley}, {Tielens}, \&
  {Millar}}]{charnley92}
{Charnley}, S.~B., {Tielens}, A.~G.~G.~M., \& {Millar}, T.~J. 1992, \apjl, 399,
  L71, \dodoi{10.1086/186609}

\bibitem[{{Chuang} {et~al.}(2016){Chuang}, {Fedoseev}, {Ioppolo}, {van
  Dishoeck}, \& {Linnartz}}]{chuang16}
{Chuang}, K.-J., {Fedoseev}, G., {Ioppolo}, S., {van Dishoeck}, E.~F., \&
  {Linnartz}, H. 2016, \mnras, 455, 1702, \dodoi{10.1093/mnras/stv2288}

\bibitem[{{Collings} {et~al.}(2004){Collings}, {Anderson}, {Chen}, {Dever},
  {Viti}, {Williams}, \& {McCoustra}}]{collings04}
{Collings}, M.~P., {Anderson}, M.~A., {Chen}, R., {et~al.} 2004, \mnras, 354,
  1133, \dodoi{10.1111/j.1365-2966.2004.08272.x}

\bibitem[{{Crimier} {et~al.}(2010){Crimier}, {Ceccarelli}, {Maret},
  {Bottinelli}, {Caux}, {Kahane}, {Lis}, \& {Olofsson}}]{crimier10}
{Crimier}, N., {Ceccarelli}, C., {Maret}, S., {et~al.} 2010, \aap, 519, A65,
  \dodoi{10.1051/0004-6361/200913112}

\bibitem[{{Cuppen} {et~al.}(2018){Cuppen}, {Fredon}, {Lamberts}, {Penteado},
  {Simons}, \& {Walsh}}]{cuppen18}
{Cuppen}, H.~M., {Fredon}, A., {Lamberts}, T., {et~al.} 2018, in IAU Symposium,
  Vol. 332, IAU Symposium, ed. M.~{Cunningham}, T.~{Millar}, \& Y.~{Aikawa},
  293--304

\bibitem[{{D'Anna} {et~al.}(2003){D'Anna}, {Bakkan}, {Beukes}, {Nielsen},
  {Brudnik}, \& {Jodkowski}}]{d'anna03}
{D'Anna}, B., {Bakkan}, V., {Beukes}, J.~A., {et~al.} 2003, PCCP, 5, 1790,
  \dodoi{10.1039/b211234p}

\bibitem[{{Enrique-Romero} {et~al.}(2016){Enrique-Romero}, {Rimola},
  {Ceccarelli}, \& {Balucani}}]{enrique-romero16}
{Enrique-Romero}, J., {Rimola}, A., {Ceccarelli}, C., \& {Balucani}, N. 2016,
  \mnras, 459, L6, \dodoi{10.1093/mnrasl/slw031}

\bibitem[{{Enrique-Romero} {et~al.}(2020){Enrique-Romero},
  {{\'A}lvarez-Barcia}, {Kolb}, {Rimola}, {Ceccarelli}, {Balucani}, {Meisner},
  {Ugliengo}, {Lamberts}, \& {K{\"a}stner}}]{enrique-romero20}
{Enrique-Romero}, J., {{\'A}lvarez-Barcia}, S., {Kolb}, F.~J., {et~al.} 2020,
  \mnras, 493, 2523, \dodoi{10.1093/mnras/staa484}

\bibitem[{{Fayolle} {et~al.}(2011){Fayolle}, {{\"O}berg}, {Cuppen}, {Visser},
  \& {Linnartz}}]{fayolle11}
{Fayolle}, E.~C., {{\"O}berg}, K.~I., {Cuppen}, H.~M., {Visser}, R., \&
  {Linnartz}, H. 2011, \aap, 529, A74, \dodoi{10.1051/0004-6361/201016121}

\bibitem[{{Fedoseev} {et~al.}(2016){Fedoseev}, {Chuang}, {van Dishoeck},
  {Ioppolo}, \& {Linnartz}}]{fedoseev16}
{Fedoseev}, G., {Chuang}, K.~J., {van Dishoeck}, E.~F., {Ioppolo}, S., \&
  {Linnartz}, H. 2016, \mnras, 460, 4297, \dodoi{10.1093/mnras/stw1028}

\bibitem[{{Froebrich}(2005)}]{froebrich05}
{Froebrich}, D. 2005, \apjs, 156, 169, \dodoi{10.1086/426441}

\bibitem[{{Furuya} \& {Aikawa}(2014)}]{furuya14}
{Furuya}, K., \& {Aikawa}, Y. 2014, \apj, 790, 97,
  \dodoi{10.1088/0004-637X/790/2/97}

\bibitem[{{Furuya} {et~al.}(2015){Furuya}, {Aikawa}, {Hincelin}, {Hassel},
  {Bergin}, {Vasyunin}, \& {Herbst}}]{furuya15}
{Furuya}, K., {Aikawa}, Y., {Hincelin}, U., {et~al.} 2015, \aap, 584, A124,
  \dodoi{10.1051/0004-6361/201527050}

\bibitem[{{Furuya} {et~al.}(2017){Furuya}, {Drozdovskaya}, {Visser}, {van
  Dishoeck}, {Walsh}, {Harsono}, {Hincelin}, \& {Taquet}}]{furuya17}
{Furuya}, K., {Drozdovskaya}, M.~N., {Visser}, R., {et~al.} 2017, \aap, 599,
  A40, \dodoi{10.1051/0004-6361/201629269}

\bibitem[{{Garrod}(2013)}]{garrod13}
{Garrod}, R.~T. 2013, \apj, 765, 60, \dodoi{10.1088/0004-637X/765/1/60}

\bibitem[{{Garrod} \& {Herbst}(2006)}]{garrod06}
{Garrod}, R.~T., \& {Herbst}, E. 2006, \aap, 457, 927,
  \dodoi{10.1051/0004-6361:20065560}

\bibitem[{{Garrod} \& {Widicus Weaver}(2013)}]{garrodCR13}
{Garrod}, R.~T., \& {Widicus Weaver}, S.~L. 2013, Chemical Reviews, 113, 8939,
  \dodoi{10.1021/cr400147g}

\bibitem[{{Gibb} {et~al.}(2004){Gibb}, {Whittet}, {Boogert}, \&
  {Tielens}}]{gibb04}
{Gibb}, E.~L., {Whittet}, D.~C.~B., {Boogert}, A.~C.~A., \& {Tielens},
  A.~G.~G.~M. 2004, \apjs, 151, 35, \dodoi{10.1086/381182}

\bibitem[{{Graninger} {et~al.}(2016){Graninger}, {Wilkins}, \&
  {{\"O}berg}}]{graninger16}
{Graninger}, D.~M., {Wilkins}, O.~H., \& {{\"O}berg}, K.~I. 2016, \apj, 819,
  140, \dodoi{10.3847/0004-637X/819/2/140}

\bibitem[{{Hasegawa} \& {Herbst}(1993)}]{hase93}
{Hasegawa}, T.~I., \& {Herbst}, E. 1993, \mnras, 263, 589,
  \dodoi{10.1093/mnras/263.3.589}

\bibitem[{{Hasegawa} {et~al.}(1992){Hasegawa}, {Herbst}, \& {Leung}}]{hase92}
{Hasegawa}, T.~I., {Herbst}, E., \& {Leung}, C.~M. 1992, \apjs, 82, 167,
  \dodoi{10.1086/191713}

\bibitem[{{Hassel} {et~al.}(2010){Hassel}, {Herbst}, \& {Bergin}}]{hassel10}
{Hassel}, G.~E., {Herbst}, E., \& {Bergin}, E.~A. 2010, \aap, 515, A66,
  \dodoi{10.1051/0004-6361/200913896}

\bibitem[{{Hassel} {et~al.}(2008){Hassel}, {Herbst}, \& {Garrod}}]{hassel08}
{Hassel}, G.~E., {Herbst}, E., \& {Garrod}, R.~T. 2008, \apj, 681, 1385,
  \dodoi{10.1086/588185}

\bibitem[{{Herbst} \& {van Dishoeck}(2009)}]{herbst09}
{Herbst}, E., \& {van Dishoeck}, E.~F. 2009, \araa, 47, 427,
  \dodoi{10.1146/annurev-astro-082708-101654}

\bibitem[{{Higuchi} {et~al.}(2018){Higuchi}, {Sakai}, {Watanabe},
  {L{\'o}pez-Sepulcre}, {Yoshida}, {Oya}, {Imai}, {Zhang}, {Ceccarelli},
  {Lefloch}, {Codella}, {Bachiller}, {Hirota}, {Sakai}, \&
  {Yamamoto}}]{higuchi18}
{Higuchi}, A.~E., {Sakai}, N., {Watanabe}, Y., {et~al.} 2018, \apjs, 236, 52,
  \dodoi{10.3847/1538-4365/aabfe9}

\bibitem[{{Imai} {et~al.}(2019){Imai}, {Oya}, {Sakai}, {L{\'o}pez-Sepulcre},
  {Watanabe}, \& {Yamamoto}}]{imai19}
{Imai}, M., {Oya}, Y., {Sakai}, N., {et~al.} 2019, \apjl, 873, L21,
  \dodoi{10.3847/2041-8213/ab0c20}

\bibitem[{{Imai} {et~al.}(2016){Imai}, {Sakai}, {Oya}, {L{\'o}pez-Sepulcre},
  {Watanabe}, {Ceccarelli}, {Lefloch}, {Caux}, {Vastel}, {Kahane}, {Sakai},
  {Hirota}, {Aikawa}, \& {Yamamoto}}]{imai16}
{Imai}, M., {Sakai}, N., {Oya}, Y., {et~al.} 2016, \apjl, 830, L37,
  \dodoi{10.3847/2041-8205/830/2/L37}

\bibitem[{{Inoue} \& {Inutsuka}(2012)}]{inoue12}
{Inoue}, T., \& {Inutsuka}, S.-i. 2012, \apj, 759, 35,
  \dodoi{10.1088/0004-637X/759/1/35}

\bibitem[{{J{\o}rgensen} {et~al.}(2013){J{\o}rgensen}, {Visser}, {Sakai},
  {Bergin}, {Brinch}, {Harsono}, {Lindberg}, {van Dishoeck}, {Yamamoto},
  {Bisschop}, \& {Persson}}]{jorgensen13}
{J{\o}rgensen}, J.~K., {Visser}, R., {Sakai}, N., {et~al.} 2013, \apjl, 779,
  L22, \dodoi{10.1088/2041-8205/779/2/L22}

\bibitem[{{J{\o}rgensen} {et~al.}(2016){J{\o}rgensen}, {van der Wiel},
  {Coutens}, {Lykke}, {M{\"u}ller}, {van Dishoeck}, {Calcutt}, {Bjerkeli},
  {Bourke}, {Drozdovskaya}, {Favre}, {Fayolle}, {Garrod}, {Jacobsen},
  {{\"O}berg}, {Persson}, \& {Wampfler}}]{jorgensen16}
{J{\o}rgensen}, J.~K., {van der Wiel}, M.~H.~D., {Coutens}, A., {et~al.} 2016,
  \aap, 595, A117, \dodoi{10.1051/0004-6361/201628648}

\bibitem[{{Karska} {et~al.}(2013){Karska}, {Herczeg}, {van Dishoeck},
  {Wampfler}, {Kristensen}, {Goicoechea}, {Visser}, {Nisini}, {San
  Jos{\'e}-Garc{\'\i}a}, {Bruderer}, {{\'S}niady}, {Doty}, {Fedele},
  {Y{\i}ld{\i}z}, {Benz}, {Bergin}, {Caselli}, {Herpin}, {Hogerheijde},
  {Johnstone}, {J{\o}rgensen}, {Liseau}, {Tafalla}, {van der Tak}, \&
  {Wyrowski}}]{karska13}
{Karska}, A., {Herczeg}, G.~J., {van Dishoeck}, E.~F., {et~al.} 2013, \aap,
  552, A141, \dodoi{10.1051/0004-6361/201220028}

\bibitem[{{Kayanuma} {et~al.}(2019){Kayanuma}, {Shoji}, {Furuya}, {Kamiya},
  {Aikawa}, {Umemura}, \& {Shigeta}}]{kayanuma19}
{Kayanuma}, M., {Shoji}, M., {Furuya}, K., {et~al.} 2019, J. Phys. Chem. A,
  123, 5633, \dodoi{10.1021/acs.jpca.9b02345}

\bibitem[{{Kristensen} {et~al.}(2012){Kristensen}, {van Dishoeck}, {Bergin},
  {Visser}, {Y{\i}ld{\i}z}, {San Jose-Garcia}, {J{\o}rgensen}, {Herczeg},
  {Johnstone}, {Wampfler}, {Benz}, {Bruderer}, {Cabrit}, {Caselli}, {Doty},
  {Harsono}, {Herpin}, {Hogerheijde}, {Karska}, {van Kempen}, {Liseau},
  {Nisini}, {Tafalla}, {van der Tak}, \& {Wyrowski}}]{kristensen12}
{Kristensen}, L.~E., {van Dishoeck}, E.~F., {Bergin}, E.~A., {et~al.} 2012,
  \aap, 542, A8, \dodoi{10.1051/0004-6361/201118146}

\bibitem[{{Lindberg} {et~al.}(2015){Lindberg}, {J{\o}rgensen}, {Watanabe},
  {Bisschop}, {Sakai}, \& {Yamamoto}}]{lindberg15}
{Lindberg}, J.~E., {J{\o}rgensen}, J.~K., {Watanabe}, Y., {et~al.} 2015, \aap,
  584, A28, \dodoi{10.1051/0004-6361/201526222}

\bibitem[{{L{\'o}pez-Sepulcre} {et~al.}(2017){L{\'o}pez-Sepulcre}, {Sakai},
  {Neri}, {Imai}, {Oya}, {Ceccarelli}, {Higuchi}, {Aikawa}, {Bottinelli},
  {Caux}, {Hirota}, {Kahane}, {Lefloch}, {Vastel}, {Watanabe}, \&
  {Yamamoto}}]{lopez17}
{L{\'o}pez-Sepulcre}, A., {Sakai}, N., {Neri}, R., {et~al.} 2017, \aap, 606,
  A121, \dodoi{10.1051/0004-6361/201630334}

\bibitem[{{Lu} {et~al.}(2018){Lu}, {Chang}, \& {Aikawa}}]{lu18}
{Lu}, Y., {Chang}, Q., \& {Aikawa}, Y. 2018, \apj, 869, 165,
  \dodoi{10.3847/1538-4357/aaeed8}

\bibitem[{{Majumdar} {et~al.}(2017){Majumdar}, {Gratier}, {Andron}, {Wakelam},
  \& {Caux}}]{majumdar17}
{Majumdar}, L., {Gratier}, P., {Andron}, I., {Wakelam}, V., \& {Caux}, E. 2017,
  \mnras, 467, 3525, \dodoi{10.1093/mnras/stx259}

\bibitem[{{Masunaga} \& {Inutsuka}(2000)}]{masunaga00}
{Masunaga}, H., \& {Inutsuka}, S.-i. 2000, \apj, 531, 350,
  \dodoi{10.1086/308439}

\bibitem[{{Masunaga} {et~al.}(1998){Masunaga}, {Miyama}, \&
  {Inutsuka}}]{masunaga98}
{Masunaga}, H., {Miyama}, S.~M., \& {Inutsuka}, S.-i. 1998, \apj, 495, 346,
  \dodoi{10.1086/305281}

\bibitem[{{Murillo} {et~al.}(2018){Murillo}, {van Dishoeck}, {van der Wiel},
  {J{\o}rgensen}, {Drozdovskaya}, {Calcutt}, \& {Harsono}}]{murillo18}
{Murillo}, N.~M., {van Dishoeck}, E.~F., {van der Wiel}, M.~H.~D., {et~al.}
  2018, \aap, 617, A120, \dodoi{10.1051/0004-6361/201731724}

\bibitem[{{Noble} {et~al.}(2015){Noble}, {Theule}, {Congiu}, {Dulieu},
  {Bonnin}, {Bassas}, {Duvernay}, {Danger}, \& {Chiavassa}}]{noble15}
{Noble}, J.~A., {Theule}, P., {Congiu}, E., {et~al.} 2015, \aap, 576, A91,
  \dodoi{10.1051/0004-6361/201425403}

\bibitem[{{Nomura} {et~al.}(2009){Nomura}, {Aikawa}, {Nakagawa}, \&
  {Millar}}]{nomura09}
{Nomura}, H., {Aikawa}, Y., {Nakagawa}, Y., \& {Millar}, T.~J. 2009, \aap, 495,
  183, \dodoi{10.1051/0004-6361:200810206}

\bibitem[{{{\"O}berg} {et~al.}(2008){{\"O}berg}, {Boogert}, {Pontoppidan},
  {Blake}, {Evans}, {Lahuis}, \& {van Dishoeck}}]{oberg08}
{{\"O}berg}, K.~I., {Boogert}, A.~C.~A., {Pontoppidan}, K.~M., {et~al.} 2008,
  \apj, 678, 1032, \dodoi{10.1086/533432}

\bibitem[{{{\"O}berg} {et~al.}(2011){{\"O}berg}, {Boogert}, {Pontoppidan}, {van
  den Broek}, {van Dishoeck}, {Bottinelli}, {Blake}, \& {Evans}}]{oberg11}
---. 2011, \apj, 740, 109, \dodoi{10.1088/0004-637X/740/2/109}

\bibitem[{{Oya} {et~al.}(2016){Oya}, {Sakai}, {L{\'o}pez-Sepulcre}, {Watanabe},
  {Ceccarelli}, {Lefloch}, {Favre}, \& {Yamamoto}}]{oya16}
{Oya}, Y., {Sakai}, N., {L{\'o}pez-Sepulcre}, A., {et~al.} 2016, \apj, 824, 88,
  \dodoi{10.3847/0004-637X/824/2/88}

\bibitem[{{Oya} {et~al.}(2017){Oya}, {Sakai}, {Watanabe}, {Higuchi}, {Hirota},
  {L{\'o}pez-Sepulcre}, {Sakai}, {Aikawa}, {Ceccarelli}, {Lefloch}, {Caux},
  {Vastel}, {Kahane}, \& {Yamamoto}}]{oya17}
{Oya}, Y., {Sakai}, N., {Watanabe}, Y., {et~al.} 2017, \apj, 837, 174,
  \dodoi{10.3847/1538-4357/aa6300}

\bibitem[{{Oya} {et~al.}(2019){Oya}, {L{\'o}pez-Sepulcre}, {Sakai}, {Watanabe},
  {Higuchi}, {Hirota}, {Aikawa}, {Sakai}, {Ceccarelli}, {Lefloch}, {Caux},
  {Vastel}, {Kahane}, \& {Yamamoto}}]{oya19}
{Oya}, Y., {L{\'o}pez-Sepulcre}, A., {Sakai}, N., {et~al.} 2019, \apj, 881,
  112, \dodoi{10.3847/1538-4357/ab2b97}

\bibitem[{{Qasim} {et~al.}(2018){Qasim}, {Chuang}, {Fedoseev}, {Ioppolo},
  {Boogert}, \& {Linnartz}}]{qasim18}
{Qasim}, D., {Chuang}, K.-J., {Fedoseev}, G., {et~al.} 2018, \aap, 612, A83,
  \dodoi{10.1051/0004-6361/201732355}

\bibitem[{{Remijan} \& {Hollis}(2006)}]{remijan06}
{Remijan}, A.~J., \& {Hollis}, J.~M. 2006, \apj, 640, 842,
  \dodoi{10.1086/500239}

\bibitem[{{Rodgers} \& {Charnley}(2003)}]{rodgers03}
{Rodgers}, S.~D., \& {Charnley}, S.~B. 2003, \apj, 585, 355,
  \dodoi{10.1086/345497}

\bibitem[{{Ruaud} {et~al.}(2015){Ruaud}, {Loison}, {Hickson}, {Gratier},
  {Hersant}, \& {Wakelam}}]{ruaud15}
{Ruaud}, M., {Loison}, J.~C., {Hickson}, K.~M., {et~al.} 2015, \mnras, 447,
  4004, \dodoi{10.1093/mnras/stu2709}

\bibitem[{{Ruaud} {et~al.}(2016){Ruaud}, {Wakelam}, \& {Hersant}}]{ruaud16}
{Ruaud}, M., {Wakelam}, V., \& {Hersant}, F. 2016, \mnras, 459, 3756,
  \dodoi{10.1093/mnras/stw887}

\bibitem[{{Sakai} {et~al.}(2009{\natexlab{a}}){Sakai}, {Sakai}, {Hirota},
  {Burton}, \& {Yamamoto}}]{sakai09a}
{Sakai}, N., {Sakai}, T., {Hirota}, T., {Burton}, M., \& {Yamamoto}, S.
  2009{\natexlab{a}}, \apj, 697, 769, \dodoi{10.1088/0004-637X/697/1/769}

\bibitem[{{Sakai} {et~al.}(2008){Sakai}, {Sakai}, {Hirota}, \&
  {Yamamoto}}]{sakai08}
{Sakai}, N., {Sakai}, T., {Hirota}, T., \& {Yamamoto}, S. 2008, \apj, 672, 371,
  \dodoi{10.1086/523635}

\bibitem[{{Sakai} {et~al.}(2009{\natexlab{b}}){Sakai}, {Sakai}, {Hirota}, \&
  {Yamamoto}}]{sakai09b}
---. 2009{\natexlab{b}}, \apj, 702, 1025, \dodoi{10.1088/0004-637X/702/2/1025}

\bibitem[{{Sakai} \& {Yamamoto}(2013)}]{sakai13}
{Sakai}, N., \& {Yamamoto}, S. 2013, Chemical Reviews, 113, 8981,
  \dodoi{10.1021/cr4001308}

\bibitem[{{Sewi{\l}o} {et~al.}(2018){Sewi{\l}o}, {Indebetouw}, {Charnley},
  {Zahorecz}, {Oliveira}, {van Loon}, {Ward}, {Chen}, {Wiseman}, {Fukui},
  {Kawamura}, {Meixner}, {Onishi}, \& {Schilke}}]{sewilo18}
{Sewi{\l}o}, M., {Indebetouw}, R., {Charnley}, S.~B., {et~al.} 2018, \apjl,
  853, L19, \dodoi{10.3847/2041-8213/aaa079}

\bibitem[{{Shimonishi} {et~al.}(2020){Shimonishi}, {Das}, {Sakai}, {Tanaka},
  {Aikawa}, {Onaka}, {Watanabe}, \& {Nishimura}}]{shimonishi20}
{Shimonishi}, T., {Das}, A., {Sakai}, N., {et~al.} 2020, \apj, 891, 164,
  \dodoi{10.3847/1538-4357/ab6e6b}

\bibitem[{{Shimonishi} {et~al.}(2016){Shimonishi}, {Onaka}, {Kawamura}, \&
  {Aikawa}}]{shimonishi16}
{Shimonishi}, T., {Onaka}, T., {Kawamura}, A., \& {Aikawa}, Y. 2016, \apj, 827,
  72, \dodoi{10.3847/0004-637X/827/1/72}

\bibitem[{{Shimonishi} {et~al.}(2018){Shimonishi}, {Watanabe}, {Nishimura},
  {Aikawa}, {Yamamoto}, {Onaka}, {Sakai}, \& {Kawamura}}]{shimonishi18}
{Shimonishi}, T., {Watanabe}, Y., {Nishimura}, Y., {et~al.} 2018, \apj, 862,
  102, \dodoi{10.3847/1538-4357/aacd0c}

\bibitem[{{Sivakumaran} {et~al.}(2003){Sivakumaran}, {H\''olscher}, {Dillon},
  \& {Crowley}}]{sivakumaran03}
{Sivakumaran}, V., {H\''olscher}, D., {Dillon}, T.~J., \& {Crowley}, J.~N.
  2003, PCCP, 5, 4821, \dodoi{10.1039/b306859e}

\bibitem[{{Snow} \& {McCall}(2006)}]{snow06}
{Snow}, T., \& {McCall}, B. 2006, Annual Review of Astronomy and Astrophysics,
  44, 367, \dodoi{10.1146/annurev.astro.43.072103.150624}

\bibitem[{{Spezzano} {et~al.}(2016){Spezzano}, {Bizzocchi}, {Caselli}, {Harju},
  \& {Br{\"u}nken}}]{spezzano16}
{Spezzano}, S., {Bizzocchi}, L., {Caselli}, P., {Harju}, J., \& {Br{\"u}nken},
  S. 2016, \aap, 592, L11, \dodoi{10.1051/0004-6361/201628652}

\bibitem[{{Suzuki} {et~al.}(1992){Suzuki}, {Yamamoto}, {Ohishi}, {Kaifu},
  {Ishikawa}, {Hirahara}, \& {Takano}}]{suzuki92}
{Suzuki}, H., {Yamamoto}, S., {Ohishi}, M., {et~al.} 1992, \apj, 392, 551,
  \dodoi{10.1086/171456}

\bibitem[{{Taquet} {et~al.}(2014){Taquet}, {Charnley}, \&
  {Sipil{\"a}}}]{taquet14}
{Taquet}, V., {Charnley}, S.~B., \& {Sipil{\"a}}, O. 2014, \apj, 791, 1,
  \dodoi{10.1088/0004-637X/791/1/1}

\bibitem[{{Taquet} {et~al.}(2015){Taquet}, {L{\'o}pez-Sepulcre}, {Ceccarelli},
  {Neri}, {Kahane}, \& {Charnley}}]{taquet15}
{Taquet}, V., {L{\'o}pez-Sepulcre}, A., {Ceccarelli}, C., {et~al.} 2015, \apj,
  804, 81, \dodoi{10.1088/0004-637X/804/2/81}

\bibitem[{{Taquet} {et~al.}(2016){Taquet}, {Wirstr{\"o}m}, \&
  {Charnley}}]{taquet16}
{Taquet}, V., {Wirstr{\"o}m}, E.~S., \& {Charnley}, S.~B. 2016, \apj, 821, 46,
  \dodoi{10.3847/0004-637X/821/1/46}

\bibitem[{{Terebey} {et~al.}(1984){Terebey}, {Shu}, \& {Cassen}}]{terebey84}
{Terebey}, S., {Shu}, F.~H., \& {Cassen}, P. 1984, \apj, 286, 529,
  \dodoi{10.1086/162628}

\bibitem[{{van Dishoeck} {et~al.}(1995){van Dishoeck}, {Blake}, {Jansen}, \&
  {Groesbeck}}]{vandishoeck95}
{van Dishoeck}, E.~F., {Blake}, G.~A., {Jansen}, D.~J., \& {Groesbeck}, T.~D.
  1995, \apj, 447, 760, \dodoi{10.1086/175915}

\bibitem[{{Vasyunin} \& {Herbst}(2013)}]{vasyunin13}
{Vasyunin}, A.~I., \& {Herbst}, E. 2013, \apj, 762, 86,
  \dodoi{10.1088/0004-637X/762/2/86}

\bibitem[{{Vaytet} \& {Haugb{\o}lle}(2017)}]{vaytet17}
{Vaytet}, N., \& {Haugb{\o}lle}, T. 2017, \aap, 598, A116,
  \dodoi{10.1051/0004-6361/201628194}

\bibitem[{{Vidal} {et~al.}(2019){Vidal}, {Gratier}, {Vaytet}, {Coutens}, \&
  {Wakelam}}]{vidal19}
{Vidal}, T. H.~G., {Gratier}, P., {Vaytet}, N., {Coutens}, A., \& {Wakelam}, V.
  2019, \mnras, 486, 5197, \dodoi{10.1093/mnras/stz1214}

\bibitem[{Whittet {et~al.}(2009)Whittet, Cook, Chiar, Pendleton, Shenoy, \&
  Gerakines}]{whittet09}
Whittet, D. C.~B., Cook, A.~M., Chiar, J.~E., {et~al.} 2009, The Astrophysical
  Journal, 695, 94, \dodoi{10.1088/0004-637x/695/1/94}

\bibitem[{{Xu} {et~al.}(2006){Xu}, {Zhu}, \& {Lin}}]{xu06}
{Xu}, S., {Zhu}, R.~S., \& {Lin}, M.~C. 2006, International Journal of Chemical
  Kinetics

\bibitem[{{Yoshida} {et~al.}(2019){Yoshida}, {Sakai}, {Nishimura}, {Tokudome},
  {Watanabe}, {Sakai}, {Takano}, \& {Yamamoto}}]{yoshida19}
{Yoshida}, K., {Sakai}, N., {Nishimura}, Y., {et~al.} 2019, \pasj, 71, S18,
  \dodoi{10.1093/pasj/psy136}

\end{thebibliography}

\end{document}